\documentclass[12pt]{article}
\input xy
\xyoption{all}
\input epsf.tex
\usepackage{pstricks}
\usepackage{psfrag}

\usepackage{graphicx}
\usepackage[small,loose]{subfigure}
\usepackage{mathrsfs}
\usepackage{bbm} 
\usepackage{cancel}
\usepackage{amssymb,amsmath}
\usepackage{fleqn}

\def\harr#1#2{\smash{\mathop{\hbox to .3in{\rightarrowfill}}
 \limits^{\scriptstyle#1}_{\scriptstyle#2}}}

\def\s2{\frac{1}{\sqrt2}}

\def\be{\begin{equation}}
\def\ee{\end{equation}}
\def\beqa{\begin{eqnarray}}
\def\eeqa{\end{eqnarray}}

\def\Tr{{\rm Tr \,}}

\def\Dsl{\,\raise.15ex\hbox{/}\mkern-13.5mu D} 
\def\d3{d^3}

\def\IR{\mathbb{R}}
\def\IC{\mathbb{C}}
\def\IZ{\mathbb{Z}}
\def\ItZ{\ensuremath{\widetilde{\mathbb{Z}}}}

\def\IS{\ensuremath{\mathbb{S}}}

\def\IRP{\ensuremath{\mathbb{R}{\rm P}}}

\newcommand{\drawsquare}[2]{\hbox{%
\rule{#2pt}{#1pt}\hskip-#2pt
\rule{#1pt}{#2pt}\hskip-#1pt
\rule[#1pt]{#1pt}{#2pt}}\rule[#1pt]{#2pt}{#2pt}\hskip-#2pt
\rule{#2pt}{#1pt}}

\newcommand{\fund}{\raisebox{-.5pt}{\drawsquare{6.5}{0.4}}}
\newcommand{\Ysymm}{\raisebox{-.5pt}{\drawsquare{6.5}{0.4}}\hskip-0.4pt%
        \raisebox{-.5pt}{\drawsquare{6.5}{0.4}}}
\newcommand{\Yasymm}{\raisebox{-3.5pt}{\drawsquare{6.5}{0.4}}\hskip-6.9pt%
        \raisebox{3pt}{\drawsquare{6.5}{0.4}}}
\newcommand{\antifund}{\overline{\fund}}

\makeatletter \@addtoreset{equation}{section} \makeatother

\begin{document}

\vspace{.5cm}
\begin{center}
\Large{\bf D-Branes in Orientifolds and Orbifolds and Kasparov KK-Theory}\\
\vspace{1cm}

\large H. Garc\'{\i}a-Compe\'an$^{a,b}$, W. Herrera-Su\'arez$^{b}$,\\[2mm]
 B. A. Itz\'a-Ortiz$^{c}$, O. Loaiza-Brito$^{a}$\\[2mm]

{\small \em $^a$ Centro de Investigaci\'on y de Estudios Avanzados del IPN, Unidad Monterrey PIIT,}\\
{\small\em Autopista Monterrey-Aeropuerto Km 10, 66600 Apodaca, Nuevo Le\'on, M\'exico}\\[4mm]

{\small \em $^b$Departamento de F\'{\i}sica, Centro de Investigaci\'on y de Estudios Avanzados del IPN}\\
{\small\em P.O. Box 14-740, 07000 M\'exico D.F., M\'exico}\\[4mm]

{\small \em $^c$Centro de Investigaci\'on en Matem\'aticas,
Universidad Aut\'onoma del Estado de Hidalgo\\
Carretera Pachuca-Tulancingo Km. 4.5, Pachuca, Hidalgo 42184, M\'exico}\\[4mm]

{\small {\bf E-mail:} compean@fis.cinvestav.mx,
wherrs@fis.cinvestav.mx, itza@uaeh.edu.mx,
oloaiza@fis.cinvestav.mx}

\vspace*{2cm}
\small{\bf Abstract} \\
\end{center}

\begin{center}
\begin{minipage}[h]{14.0cm} {A classification of D-branes in Type IIB O$p^{-}$
orientifolds and orbifolds in terms of
 Real and equivariant KK-groups is given. We classify
 D-branes intersecting orientifold planes from which are recovered some special
 limits as the spectrum for D-branes on top of Type I O$p^{-}$ orientifold
 and the bivariant classification of Type I D-branes.
The gauge group and transformation properties of the low energy
effective field theory living in the corresponding unstable
D-brane system are computed by extensive use of Clifford algebras.
Some speculations about the existence of other versions of
KK-groups, based on physical insights, are proposed. In the
orbifold case, some known results concerning D-branes intersecting
orbifolds are reproduced and generalized. Finally, the gauge
theory of unstable systems in these orbifolds is recovered.}
 \end{minipage}
\end{center}

\bigskip
\bigskip

\date{\today}

\vspace{3cm}

\leftline{September 2008}

\newpage

\section{Introduction}

Topological methods in physics have always been relevant in order
to describe static stable configuration of finite energy in field
and string theories. D-branes have RR charge and they are source
of RR fields. Both of them are classified by K-theory in all
different theories. The recipe is that the K-theory is described
by the classes of pairs of gauge bundles over the worldvolume of
the D9-$\overline{\hbox{D}9}$ pair of the Type IIB string theory
or on the non-BPS D9 of the Type IIA theory. This description
classifies all lower dimensional D-branes coming from tachyon
condensation called descendent branes. However the inverse process
of constructing higher dimensional D-branes from the lowest
dimensional unstable systems of D-instantons is also possible.
This description is given by using K-homology.

The incorporation of the K-homology description in the context of
Matrix theory was done in \cite{Asakawa:2001vm}. This was called
the K-matrix theory and is based in the process involving
configurations of non-BPS instanton in Type IIA string theory and
D-instantons - anti D-instantons in Type IIB theory. From these
configurations, higher-dimensional D-branes can be constructed and
they are classified (through their worldvolumes) by the K-homology
groups. The D-branes are described and thus represented by
equivalence classes of Connes spectral triples (analytical data)
used in noncommutative geometry. This equivalence is physically
defined in terms of the gauge equivalence and the charge
conservation. For different approaches of K-homology to D-brane
classification see \cite{Periwal:2000eb, Reis:2005pp, Reis:2006th,
Szabo:2002jv}.

K-theory and K-homology are dual one of each other and it is
compelling to use the Kasparov (complex) KK-theory, which is a
generalization of both theories. This was done in
\cite{Asakawa:2001vm} where the procedure of the construction of
ascendent-descendent brane configuration was implemented on the
product space-time $X \times Y$ with the world-volume of the
unstable D-brane wrapped on $Y$. The D-branes are classified in a
natural way by the groups KK$^i(X,Y)$. In \cite{Asakawa:2002nv} it
was shown that D-branes of the Type I theory are classified by
(the real/Orthogonal) KKO$^i(X,Y)$.

Similarly to K-homology, there are several approaches for the
KK-theory application in describing D-brane physics
\cite{Brodzki:2006fi, Brodzki:2007hg, Reis:2006th}; but we will
concentrate in the approach from \cite{Asakawa:2001vm}.

Moreover, in the present paper we extend these results by showing
that orientifolds are classified by the Real KK-group KKR$^i(X,Y)$
and orbifolds by the equivariant KK-group KK$_G^i(X,Y)$. In
addition, we propose based on physical arguments, the existence of
different versions of the KK bifunctor which; as far as the
authors knowledge have not been discussed in the literature
before. For all these theories, the spectrum is correctly
obtained. We also give an application to exotic orientifolds.

This paper is organized as follows. In Sec.~2 a brief account of
the classification of D-branes through K-theory, K-homology and
KK-theory is given. Sec.~3 is devoted to describe D$d$-branes in
orientifold backgrounds by using the Real KKR-theory. In here we
find general formulas which involve the two cases $q \leq p$ and
$p \leq q$. Here $p$ is the dimension of the orientifold plane
O$p$ and $q$ is spatial dimension of the unstable D$q$-brane.
Sec.~4 analyzes the theory on the unstable D-brane using the
information provided by the Clifford algebras involved in the
definition of the KKR bifunctor. To be more specific we will
describe in some detail three important examples. The rest of the
cases is summarized in a table. Sec.~5 is devoted to make a
proposal for extending the classification of D-branes in
orientifolds to other theories such as the Type IIB with O$p^+$
(quaternionic) and the IIB with O$9^+$ (with gauge group
$USp(32)$) orientifold in the context of Kasparov KK-theory. At
the end of this section, we explain an application of our
formalism to exotic orientifolds. D-branes in orbifold
singularities with KK$_G$-theory are discussed in Sec.~6. Finally
in Sec.~7 we give our final remarks. Four appendices collect a
series of formal results about KK-theory.

\section{Classification of D-branes in orientifold planes }

\subsection{D-branes and K-theory}
In Type II superstring theories D-branes are constructed as
solitons on unstable systems either formed by pairs of
brane-antibranes or by single unstable D-branes \cite{Sen:1999mg}.
This means that any configuration of D-brane charges is realized
as a gauge field configuration on a stack of (sufficiently) many
D9-$\overline{\hbox{D}9}$ branes in Type IIB, or non-BPS D9-branes
in Type IIA by open string tachyon condensation. This was
interpreted as a way to classify D$d$-brane charges by gauge
bundles on the worldvolume of the D9-branes \cite{Witten:1998cd}.
Hence, D-brane charges turn out to be elements of a group
constructed from equivalence classes of vector bundles, namely
K-theory.

In Type IIB theory D$d$-brane charges are classified by the so
called complex K-theory group, which is valuated on the
transversal space (with respect to the unstable
system\footnote{Throughout this paper, what we refer as ``K-theory
group'' is really the reduced K-theory group of the compactified
space.}) to D$d$. In particular, the K-theory group classifying a
D$d$-brane in Type IIB is given by $KU(\IR^{9-d})$ which renders
the D$d$-brane as a soliton  constructed by the pair
D9-$\overline{\hbox{D}9}$. One can instead consider a D$d$-brane
as a soliton constructed from an unstable system formed by
D$q$-branes ($q>d$). The groups classifying the corresponding
vector bundles transversal to the D$d$-brane worldvolume, in a
nine-dimensional or $q$-dimensional unstable system, are
isomorphic as expected from Bott periodicity and are given by
$KU(\IR^{9-d})$ and $KU^{q-1}(\IR^{q-d})$ respectively.

D-brane classification by K-theory is a little more elaborated
once we introduce discrete actions on the background such as
orientifolds or orbifolds. For instance, Ramond-Ramond (RR) fields
on which D-branes in Type I theory are charged, have a smaller
number of degrees of freedom due to the orientifold projection.
This reduces the gauge group on the D-brane to be orthogonal or
symplectic implying that D-branes are classified by orthogonal
K-theory groups of the corresponding transversal spaces.
Specifically, D$d$-branes in type I are classified by
$KO(\IR^{9-d})$ which points out the presence of non-BPS states
carrying discrete topological charge \cite{Schwarz:1999vu}. These
D$d$-branes can also be thought of as solitonic constructions from
unstable pairs of D9-$\overline{\hbox{D}9}$ branes on top of an
orientifold nine-plane $O9^-$. In a similar context as before, we
can try to understand the construction of Type I D-branes from
lower-dimensional unstable branes (which is justified since in
general, super Yang-Mills theories in 9+1 dimensions are
non-renormalizable). In fact, it is possible to condense open
string tachyons from a pair of D$q$-$\overline{\hbox{D}q}$ on top
of the orientifold nine-plane in order to construct D$d$-branes,
which are classified by $KO^{q-1}(\IR^{q-d})$
\cite{Bergman:2000tm}.

The situation becomes much more interesting by considering the
presence of lower dimensional orientifolds $Op^-$. Classification
of D$d$-branes in such backgrounds was given in
\cite{Gukov:1999yn} and it strongly depends on which type of
orientifold background we are taking into account. It turns out
that for an orientifold with a positive squared involution
($\tau^2=1$) and $(-1)^{F_L}=1$ (i.e., for $p=1~mod~4$) the real
K-theory group which classifies D$d$-branes is
$KR(\IR^{9-p,p-d})$,
 where
\begin{align}
 \IR^{9-p,p-d}=\left(\IR^{9-p}/\Omega \cdot {\cal
I}_{9-p}\right) \times \IR^{p-d},
\end{align}
is the transversal space to the D$d$-brane. The world-sheet
operator $\Omega$ inverts the orientation of the string while the
involution ${\cal I}_{9-p}$ maps transversal coordinates to the
orientifold $x_i$ to $-x_i$. Notice that D$d$-branes on top of an
orientifold plane $Op^-$ are obtained as well by pairs of
D9-$\overline{\hbox{D}9}$ in which the open string tachyons have
been condensed. The corresponding construction from lower (than
nine) dimensional unstable systems will be studied in the next
section.

So far we have reviewed constructions of D$d$-branes from unstable
D$q$-branes ($q>d$). This means that each element of K-theory
describes a lower-dimensional (than $q$) D$d$-brane obtained by
tachyon condensation from unstable branes generalizing the D-brane
descent relations (see \cite{Sen:1999mg} and references therein).

However, it is also possible to elucidate the above construction
from the tree-level action of an unstable brane. Such an action is
constructed in the Boundary String Field Theory (BSFT) to the
superstring \cite{Kutasov:2000aq}. For instance, the action of an
unstable D9-brane in Type IIA is given by
\begin{align}
S=T_9\int d^{10}x\left(~(\ln~2) \alpha' e^{-T^2/4}\partial^\mu
T\partial_\mu T + e^{-T^2/4}\right) \label{Eq:actionD9}
\end{align}
from which a solution for the equations of motion for the tachyon
field is
\begin{align}
T= \mu X,
\end{align}
where $\mu$ is a constant and $ X$ denotes some coordinate of the
spacetime manifold.

By substituing this kink solution into the unstable D9-brane
action, we get the action of a stable D8-brane (for $\mu
\rightarrow \infty$). The argument can be generalized to show that
from the action for $N$ non-BPS D9-branes, with $N$ large enough
\begin{align}
T(X)=\mu\sum_{i=1}^{9-d}X^i\gamma_i,
\end{align}
where $\{ \gamma_i, \gamma_j\}=2\delta_{ij}$, is also a solution
for the equations of motion, giving rise to a D$d$-brane. Notice
that this expression for the tachyon field corresponds to the
Atiyah-Bott-Shapiro (ABS) construction (see for instance
\cite{Witten:1998cd, Olsen:1999xx}), which plays a relevant role
in the classification of D-branes by K-theory.

For the case of an unstable pair of brane anti-brane, the complete
tachyon field is given by
\begin{align}
\mathbf{F}= \left( \begin{array}{cc}
0&T^\dagger\\
T&0
\end{array} \right) = \mu\sum_{i=1}^{9-d}X^i \left( \begin{array}{cc}
0&\gamma_i^\dagger\\
\gamma_i&0
\end{array} \right).
\end{align}
Hence, roughly speaking,  D$d$-branes are constructed  by tachyon
condensation from higher-dimensional unstable branes and they are
classified by the gauge bundles on their corresponding transversal
spaces.

\subsection{D-branes and K-homology}
In the context of Matrix theory it is possible to construct
D$d$-branes not from higher-dimensional unstable brane systems,
but from infinitely many lower-dimensional D-branes. The idea was
developed in \cite{Terashima:2001jc} in order to construct
commutative D-branes, which turn out to be classified by
K-homology \cite{Asakawa:2001vm} in the case where the lower
dimensional D-branes are D-instantons. The basic idea is as
follows: by taking T-duality (in the euclidean space) on the nine
spatial coordinates, the action (\ref{Eq:actionD9}) for the
non-BPS D9-brane in Type IIA, generalized to $N$ D9-branes for $N$
large enough,  gives
\begin{align}
S=T_{-1}\Tr_{N\times N}\left( e^{-T^2/4}(1-c_1[\phi_\mu,
T]^2-c_2\pi^2[\phi_\mu, \phi_\nu]^2)\right),
\end{align}
which is the action for $N$ D(-1)-branes and where the
$\phi_{\mu}$ are scalar fields representing the transverse
position as a function on the coordinates $x^{\nu}$.

The corresponding equations of motion for the tachyon field have
as a solution (provided $\mu^{2}=1/c_1$)

 $$T=\frac{2\pi\mu}{\alpha '^{1/2}} p$$

 \begin{equation}
 \qquad \qquad \qquad \phi_0= \frac{1}{2\pi\alpha '^{1/2}}x, \phi_i=0, \qquad
(i=1,\cdots, 9),
\end{equation}
where the operators $x$ and $p$ are identified with the
transversal coordinates and momentum of the non-BPS D(-1)-branes.
Plugging this tachyon kink solution (in momentum) back into the
D(-1)-branes action provides a D0-brane action, whose position is
specified by the fields $\phi_i=0$. The argument can be
generalized to construct higher-dimensional D$d$-branes in Type
IIB theory from an infinite number of
D(-1)-$\overline{\hbox{D}(-1)}$ pairs, in which case the tachyon
and scalar fields are

$$T=\mu\sum_{j=0}^{d}p_j\otimes\gamma^j,$$
\begin{equation}
\qquad \qquad \qquad \phi_i^{(1)}=\phi_i^{(2)}=\frac{1}{2\pi\alpha
'^{1/2}}x^i, \qquad (i=0,\cdots, d),
\end{equation}
with $\mu\rightarrow\infty$, and  $\gamma^j$ being the
$2^{[\frac{d}{2}]}\times 2^{[\frac{d}{2}]}$ gamma matrices in $d$
dimensions. The superindices in $\phi$ stand for the instanton
brane and antibrane, respectively \cite{Terashima:2001jc}.

It follows then, that the tachyon matrix $F$  can also be written
as

\begin{align}
 F=\mu\sum_{j=0}^dp_j\otimes\Gamma^j,
\end{align}
with
\begin{align}
\Gamma^j= \left( \begin{array}{cc}
0&\gamma^{j\dagger}\\
\gamma^j&0
\end{array} \right).
\end{align}

However, since the tachyon field $T$ (which comes from the
oriented string between the instanton brane-antibrane system), is
not projected out by GSO projection, represented by the operator
$(-1)^{F_L}= \left( \begin{array}{cc}
0&I\\
I&0
\end{array} \right)$, the tachyon matrix $F$ satisfies the self-dual condition $F=F^\dagger$. This fact plays an important role in the next the sections.

Let us however, return to the question of classification of
D$d$-branes created by brane-anti-brane instantons. As in the {\it
usual} case of tachyon condensation from higher dimensional
non-BPS D-branes, the construction of D-branes from unstable
D(-1)-branes leads to their classification in terms of the so
called K-homology $K_n(X)$ \cite{Asakawa:2001vm}, which roughly
speaking, is the dual to the K-theory group $K^n(X)$ in the sense
that it has a natural pairing with the K-theory group. Instead of
classifiying vector bundles on the transverse space to a
D$d$-brane as in K-theory, K-homology classifies vector bundles on
the worldvolume of the extended D$d$-branes constructed from
unstable $D(-1)$-branes \footnote{More precisely, the topological
K-homology of any locally compact space $X$ classifies triples
$(M, E, \phi),$ where
\begin{itemize}

   \item $M$ is a compact $spin^{c}$-manifold without boundary.

   \item $E$ is a complex vector bundle over $M$.


   \item $\phi\colon M \to X$ is an embedding of $M$ in $X$.

   \end{itemize}

   The equivalence relations on the triples $(M, E, \phi)$ that define the K-homology of $X$ have a nice physical
   interpretation in terms of D-brane processes. In fact the components of the triples $(M,E,\phi)$ are easily
   interpreted as the worldvolume manifold $M$ of the $D$-brane, $E$ is the
   Chan-Paton bundle on the worldvolume $M$ of the $D$-brane and $\phi$ is the embedding of the $D$-brane worldvolume
   in the ambient spacetime $X$. For more details
   see \cite{Asakawa:2001vm, Reis:2005pp}.}. This is
generalized to construct a D$d$-brane from an unstable D$q$-brane
($q<d$) with a tachyon configuration given by
\begin{align}
F=\mu\sum_{j=q+1}^{d+q}p_j\otimes\Gamma^j.
\end{align}

\subsection{Kasparov KK-theory}

By virtue of the material revisited so far, it is then natural to
combine the above two setups in order to construct a D$d$-brane by
a kind of combination of tachyon condensation from higher- and
lower-dimensional D-branes. For branes in Type II theories, the
extension was given in \cite{Asakawa:2001vm}, together with a
proposal to classify them.

In this scenario, a D$d$-brane located in coordinates $x^0,\cdots,
x^{q-s}, x^{q+1}, \cdots, x^{d+s}$ is constructed roughly speaking
by tachyon condensation from an unstable D$q$-brane located in
coordinates $x^0,\cdots ,x^q$ with a tachyon configuration given
by
\begin{align}
F=\mu\sum_{i=0}^{s}X^i\otimes\Gamma_i+\mu\sum_{j=q+1}^{d+q}p_j\otimes\Gamma^j.
\label{fulltachyon}
\end{align}

The ``part'' of the D$d$-brane localized inside the unstable
D$q$-brane is constructed by tachyon condensation as in Sen's
descent relations, while the rest can be seen as constructed from
unstable D$q$-branes as in section 2.2.

It turns out that the relevant group which classifies D$d$-branes
constructed as in the above configuration is Kasparov KK-theory
\cite{kaspa:1981wl,Asakawa:2001vm}. Let us first of all briefly
summarize some important aspects about KK-theory (see Apendix A
for a more formal and detailed description).

 KK-theory is a generalization of both K-theory and K-homology, in the sense that
   while both K-theory and K-homology are
   functors from the category of locally compact Hausdorff topological spaces
   to the category of abelian groups, (i.e. classify classes of vector bundles on the transverse space and in the worldvolume of a D-brane, respectively), KK-theory is a
   bifunctor between these categories \footnote{In fact, all the K-functors mentioned above have as domain the full
    category of $C^{*}$-algebras which includes the category of locally
    compactHausdorff spaces as a subcategory by assigning to each locally compact Hausdorff space $X$ the $C^{*}$-algebra
     of continuous $\mathbb{C}$-valued functions on $X$ vanishing at infinity. Moreover, it can be shown
     that each commutative $C^{*}$-algebra is of this form, $X$ being the space of characters of the
     algebra.} \cite{kaspa:1981wl, Blackadar:1998oa, Knudsen:1991ek, Schroder:2001ms}. The bifunctor assigns to each
   pair $(X,Y)$ of locally compact topological spaces some abelian
   group denoted $KK^{-n}(X,Y)$ for any integer $n$. Here $X$ denotes the part of the worldvolume of the D-brane (extended outside the D$q$-brane system)
   created from lower dimensional branes and $Y$ is the worldvolume of the unstable D$q$-brane from which a  D-brane is created by
   tachyon condensation (as in the descent relations). Given such identification of the topological spaces $X$ and $Y$, it is then
    expected to get some relations between KK-theory and both K-theory and K-homology groups. Indeed, if $X=\{pt\}$,
     it means we do not have a D-brane (extended in the transverse space of the D$q$-brane system) created  from lower dimensional
      branes. This implies that  D$d$-branes are entirely classified by K-theory. Then
\begin{align}
KK^{-n}(pt, Y)=K^{-n}(Y).
\label{ktheory1}
\end{align}

Similarly, for a brane fully extended outside the unstable
D$q$-brane from which it was constructed (via condensation of a
tachyon field as in Eq.(\ref{fulltachyon})), the space $Y$ is the
point-space implying that
\begin{align}
KK^{-n}(X,pt)=K_n(X).
\end{align}

Now, as in the case of K-theory which is the set of equivalence
classes of vector bundles, KK-theory is the set of equivalence
classes of Kasparov triplets $({\cal H}, \phi, T)$. In pedestrian
words, ${\cal H}$ is the set of all Chan-Paton gauge fields living
on the worldvolume of the unstable D$q$-brane ($\phi$ and $T$ are
as usual the transversal position and tachyon fields). In this
sense, a zero class representing the vacuum is gathered by a
tachyon field $T$ which condensates trivially (i.e., without a
kink solution in momentum or spatial configurations) implying that
$T^2=1$ ($T$ has been normalized) and $T$ and $\phi$ depending on
non-conjugate position and momentum, i.e. $[T,\phi]=0$. For the
case in which the tachyon condensates in a non-trivial way, it is
said that the triplet is non-trivial, representing a D-brane
configuration in which the tachyon field configuration is given by
Eq.(\ref{fulltachyon}). Hence, KK-theory is the set of triplets
which are equivalent up to the addition of a zero-class triplet.
It is, as in the case of K-theory, an equivalence which preserves
the RR charge. A formal presentation of KK-theory groups is given
in Appendix A. However, for a more detailed explanation about the
interpretation of the elements defining the Kasparov modules and
the equivalence relations involved in the definition of the
KK-groups, the reader is referred to \cite{Asakawa:2001vm}, in
which a detailed discussion on some subtleties in the choice of
the spacetime and the tachyon in the Kasparov modules is
considered.

\subsubsection{D-branes and KK-theory}
Let us consider the simple case of a D$d$-brane in Type IIB(A)
string theory, constructed from unstable D$q$-branes. In
particular, for a configuration of a D$d$-brane located in
coordinates $x^0, \cdots, x^{q-s}$ $, x^{q+1},\cdots ,x^{d+s}$,
the spaces $X$ and $Y'$ are given by $\IR^{d-q+s}$ and
$\IR^{9-q+s}$ from which the relevant KK-theory group is given
by\footnote{In (\ref{ktheory1}) we interpreted $Y$ as the
worldvolume of the unstable D$q$-brane for some integer $n$; but
similar to K-theory, $Y$ is actually the transverse space (with
respect to the D$q$-brane) of the part of the D$d$-brane localized
inside the D$q$-brane. This is achieved by making use of the Atiyah-Bott-Shapiro construction in K-theory. 
Moreover a similar meaning is assigned to
$Y'$, i.e.
is the transverse space of the part of the D$d$-brane extended
inside the unstable D$q$-brane system, but in this case the
transverse space is relative to an unstable D9 system and
consequently $n$ in (\ref{ktheory1}) is changed
depending on the string theory we are dealing with. The KK-theory
prescription in terms of $Y$ and $Y'$ are equivalent as will be
shown through out this paper.}
\begin{align}
KK^{0(-1)}(\IR^{d-q+s},\IR^{9-q+s})=K^{0(-1)}(\IR^{9-d}).
\label{kk1}
\end{align}
It is important to stress out that, as mentioned in Appendix B, it
is possible to extract information of the system through the
relation with complexified Clifford algebras $\IC l^n$ given by
\begin{align}
KK^{-n}(X,Y)=KK(C_0(X),C_0(Y)\otimes \IC l^n),
\end{align}
where $C_0(X)$ ($C_0(Y)$) denotes the algebra of complex valued
(real valued when dealing with orthogonal KK-groups) continuous
functions in $X$ ($Y$) vanishing at infinity. Such relation with
Clifford algebras shall become very important in our description
of D$d$-branes in more general backgrounds.

The next natural step is to classify D$d$-branes in Type I theory,
i.e., in the presence of a negative RR charged orientifold
nine-plane $O9^-$. This was done in \cite{Asakawa:2002nv}, where
the authors proposed that the relevant group for such
classification is the real Kasparov bifunctor, denoted as
$KKO(X,Y)$, in which roughly speaking, all complex fields become
real by the orientifold nine projection (for a formal description
and for more details, see Appendix B).

Let us consider the D$d$-brane in an $O9^-$-plane background
extended again in the coordinates $x^0, \cdots, x^{q-s}$ $,
x^{q+1},\cdots ,x^{d+s}$. In this situation the Kasparov KK-theory
group turns out to be orthogonal (real) given by
$KKO(\IR^{d-q+s},\IR^{9-q+s})$. Using the isomorphisms from
Eq.(\ref{kkbp}),  the above group reduces to
\begin{align}
KKO^{q-1}(\IR^{d+s-q},\IR^{s})=KO^{q-1}(\IR^{q-d})=KO(\IR^{9-d}),
\label{kk2}
\end{align}
as expected \cite{Bergman:2000tm}. The relation with real Clifford
algebras $Cl^{\ast,\ast}$ is given in a similar context as in Type
II
\begin{align}
KKO^{q-1}(X,Y)=KKO(C_{0}(X),C_{0}(Y) \otimes \textsl{Cl}^{1,q}),
\end{align}
for which the tachyon configuration reads
\begin{align}
F=u \sum_{\alpha = q-s+1}^{q} x_{\alpha} \otimes \Gamma^{\alpha} +
u \sum_{\beta=q+1}^{d+s} (-i\partial_{\beta}) \otimes
\Gamma^{\beta}, \label{tachyon3}
\end{align}
where $\Gamma^{\alpha}$ and $-i\Gamma^{\beta}$ are in
$M_{n}(\mathbb{R}) \otimes \textsl{Cl}^{1,q}_{odd}$ for some $n$
(see apendix B.2), satisfying

\begin{eqnarray}
 \Gamma^{\alpha \dag}=\Gamma^{\alpha},  \qquad \qquad
(-i\Gamma^{\beta})^\dag= i\Gamma^{\beta},  \qquad \qquad \qquad \qquad\\
 \{\Gamma^{\alpha}, \Gamma^{\alpha^{'}} \}=2   \nonumber
\delta^{\alpha,\alpha^{'}}, \qquad \qquad \{ \Gamma^{\beta},
\Gamma^{\beta^{'}} \}=2 \delta^{\beta,\beta^{'}}, \qquad \qquad
\{\Gamma^{\alpha},\Gamma^{\beta}\}=0. \label{gamma}
\end{eqnarray}


One can note that many physical properties of D-branes are
obtained through the analysis of Clifford algebras. Indeed,  as it
has been carefully studied in \cite{Asakawa:2002nv} for $q=9$ and
$s=9-d$ and for $q=-1$, it is possible to extract some information
as the tension of the Type I $D$-branes from the Type IIB ones and
the gauge field representations of the tachyon associated to the
worldvolume field theory of the Type I D$d$-branes constructed
from instantons. This is achieved by looking at the representation
theory of the real Clifford algebras involved in the definition of
the $KKO$-groups.

\section{D$d$-branes in orientifold  backgrounds and Real KKR-theory}

Up to now, we have reviewed the classification of D-branes in
terms of K-theory, K-homology and KK-theory. For instance, we have
seen that the Real K-theory group KR is the correct one to
classify D-branes constructed from non-BPS D9-branes in Type II
orientifolds $O1, O5$ and $O9$. On the other hand, we have a
classification of D-branes, constructed from non-BPS D$q$-branes
in Type I theory. The next thing to do is to classify D$d$-branes
by KK-theory in a more general orientifold background.

By considering only $Op^-$-planes with $p=1~mod~4$,  we shall
propose in this section that Real KK-theory\footnote{We adopt the
convention in mathematical literature by refering to the
orthogonal KK-theory as ``real'', and to the complex (with
involution) one as ``Real''.} is the correct group to classify
D$d$-branes in such backgrounds. Following closely
\cite{Asakawa:2002nv}, we shall show that our proposal can also
reproduce some of the expected properties of non-BPS and BPS
branes by studying the related
Clifford Algebra.\\

\subsection{D$d$-branes from unstable D$q$-branes in orientifold backgrounds and K-theory}
In order to know how to construct KKR-theory groups, let us first construct a K-theory group which classifies
D$d$-branes on top of an orientifold plane. Here we do not consider the case in which (part of) the D$d$-brane is constructed from lower dimensional D-branes.
As far as
we know, this group has not been reported in the literature.
However, its construction is straightforward as we shall see.

The general situation can be divided in two diferent
configurations: 1) The O$p$-plane is immersed in the unstable
D$q$-brane, i.e., $q \geqslant p$ and 2) the opposite situation in
which $p \geqslant q$. We concentrate on those cases in which the
D$d$-brane is totally immersed in the orientifold plane. More
general cases are taken into account in the KK-theory formalism.

\begin{figure}[t]
\begin{center}
\centering \epsfysize=8cm \leavevmode
\epsfbox{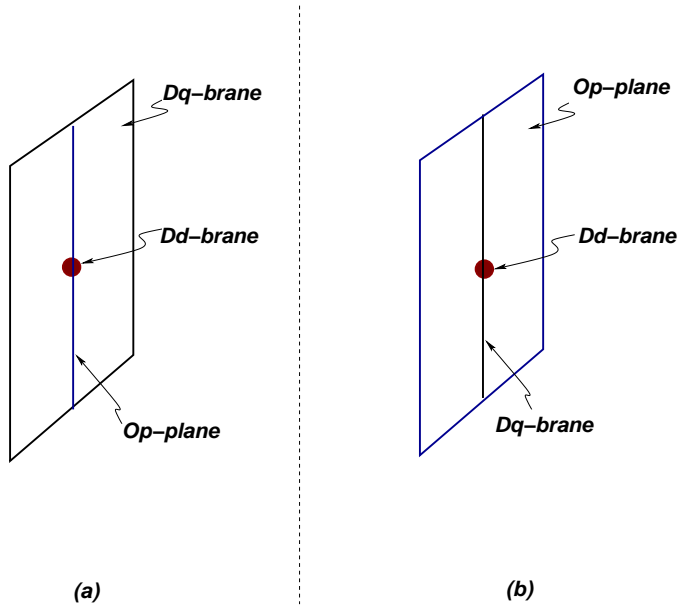}
\end{center}
\caption[]{\small \it A D$d$-brane constructed from a D$q$-brane
where $d<q$ and (a) D$q$ is dimensionally higher than the
orientifold plane $Op$, (b) $q<p$.} \label{Fig:p<q}
\end{figure}

{\bf Case 1: $q \geqslant p$}. The important issue is to construct
the transversal space to the D$d$-brane as depicted in Fig
\ref{Fig:p<q}(a). It is easy to see that such space is given by
$\IR^{(9-q)+(q-p),p-d}$, from which we can construct the
associated K-theory group as $KR(\IR^{(9-q)+(q-p),p-d})$. By using
the following relations for KR
\begin{align}
KR(\IR^{0,m})= KO(\IS^m),\nonumber\\
KR(\IR^{n,m})= KR^{n,0}(\IR^{0,m})= KR^{0,m}(\IR^{n,0}),\nonumber\\
KR^{n,m}(X)=KR(X \times \IR^{n,m}), \nonumber\\
 KR^{n,m}(X)= KR^{n-m}(X)= KR^{n-m\pm
8}(X).
\end{align}
we can rewrite the K-theory group as
\begin{align}
KR^{1-q}(\IR^{q-p,p-d}),
\end{align}
where $\IR^{q-p,p-d}$ is the transverse space of the D$d$-brane
respect the unstable D$q$-brane system.

{\bf Case 2: $p\geqslant q$}. Let us now consider the depicted in
Fig. \ref{Fig:p<q}(b). For the orientifold plane containing the
unstable D$q$-brane, the transversal space for the D$d$-brane is
$\IR^{9-p, (p-q)+(q-d)}$, for which the corresponding K-theory
group is
\begin{align}
KR(\IR^{9-p, p-d})= KR^{9-2p+q}(\IR^{0,q-d}),
\end{align}
where we have again used the isomorphisms for $KR$ in the left hand side. Notice that the K-theory group written in such a way, allows us to identify the space $\IR^{0,q-d}$ as the transversal one to the D$d$-brane with respect to the D$q$-brane, as in case 1.\\



Now, let us check if the above two formulae are consistent with
what we already know. Essentially we have two limits to check.
First of all, if $q=p=9$ we reproduce immediately the known
formula which classifies D$d$-branes in Type I theory, i.e.,
$KO(\IR^{9-d})$. The second limit to recover is Bergman's formula
for D$d$-branes in Type I theory, from unstable D$q$-branes. Hence
in this case, $p=9$ but different from $q$. In such a case, the
related K-theory group reads
\begin{align}
KR^{q-1}(\IR^{0,q-d})=KO^{q-1}(\IR^{q-d}),
\end{align}
which indeed validates our proposal.

\subsection{The Real KK-theory group}
We now proceed to define the KKR-theory groups relatad to the configurations so far discused.

We start by introducing the formal definition for the Real KK-theory group which we
shall apply in order to classify D$d$-branes in orientifold
backgrounds.

Real KK-theory groups are defined in terms of a Real
$C^{*}$-algebra which is just a complex $C^{*}$-algebra with an
additional antilinear involution
 ${\cal I}$ such that ${\cal I}(b_{1} b_{2})=
{\cal I}(b_{1}) {\cal I}(b_{2})$ and ${\cal I}(b^{*})=({\cal
I}(b))^{*}$, for every $b,b_1,b_2$ in the complex
$C^\ast$-algebra. Notice as well that (by definition) ${\cal
I}(i)=-i$.

Now, let A and B be trivially graded, separable and unital Real
$C^{*}$-algebras. An  \textit{even Kasparov Real} $A$-$B$-module
is defined as for the complex and orthogonal cases (see Appendices
A and B for details and notation), with the following additional
data:

\begin{itemize}

    \item An antilinear Real involution ${\cal I}$ on ${\cal H}_{B}$ with the following
    property:
    ${\cal I}(xb)={\cal I}(x) {\cal I}(b)$ and $({\cal I}(x),
    {\cal I}(y))= {\cal I}((x,y))$ for $x,y \in {\cal H}_{B}$ and
    $b \in B.$

    \item An antilinear Real involution ${\cal I}$ on $\textbf{B}({\cal
    H}_{B})=M_{2}(M(B \otimes {\cal K}))$ defined by ${\cal I}(T)(x)=
    {\cal I}(T( {\cal I}(x)))$ for $ x \in {\cal H}_{B}.$

    \item $\phi:A \rightarrow \textbf{B}({\cal H}_{B})$ is a
    $*$-homomorphism of Real $C^{*}$-algebras, i.e.
    $\phi({\cal I}(a))= {\cal I}(\phi (a)) \quad {\rm{for \; all \;}}  a \in
    A.$

\end{itemize}

The basic $KK$-group for Real $C^{*}$-algebras $A,B$ will be
denoted $KKR(A,B)$ and it is defined as the equivalence classes of
even Kasparov Real $A-B$-modules with the equivalence relations
defined as in the complex and real cases; with the additional
requirement that both, the homomorphisms $\phi$ and the operators
$T$ appearing in the Kasparov modules; as well as the unitary
operator generating the relation of unitary equivalence be
\textit{invariant} under the Real involution $({\cal I}(a)=a)$,
i.e they belong to the \textit{fixed point algebra} of the Real
algebra to which they belong.

The corresponding higher $KKR$-groups are denoted as $KKR^{-n}
\equiv KKR_{n}(A,B)$ and defined as in the real case, but using
the Real Clifford algebras   $^p \mathbb{C}^{n,m}$, with some Real
involution ${\cal I}_p$ which is determined in our case by the
orientifold $Op^-$ action on the Clifford generators \footnote{We
denote the Real Clifford algebras as $^p\mathbb{C}^{p,q}$ in order
to distinguish them from the complex Clifford algebras used in
complex $KK$-theory. Also, we denote a generic field $\psi$ under
the action of the involution ${\cal I}$ determined by the
orientifold $p$-plane as $^p\psi$.}. Hence, we denote the
involution action on an element $a$ of the Real Clifford algebra
as ${\cal I}_{9-p}(a)$.

In this way, we have \cite{kaspa:1981wl, Schroder:2001ms}
\begin{equation}
\qquad \qquad KKR_{m-n+r-s}(A,B)=KKR(A\otimes ~^p\IC l^{n,m},B
\otimes ~^p\IC l^{r,s}) \label{kkrcl}
\end{equation}


The $KKR^{n}$-groups are periodic mod 8 and $KKR(A,B)=KKO(A,B)$ if
both $A$ and $B$ have trivial Real involution
\cite{Schroder:2001ms}. A Bott periodicity result also holds for
Real $KK$-theory:

$$KKR^{k}(X,Y)=KKR^{k+m-n}(X\times\mathbb{R}^{m,n},Y)=KKR^{k-m+n}(X,Y \times\mathbb{R}^{m,n}),$$
\begin{equation}
\qquad \qquad \qquad \qquad KKR^{-m}(pt, Y)= KR^{-m}(Y).
\label{isoKKR}
\end{equation}


One important example that will be useful in Sec. 4 is $Y=pt$. In
terms of the Kasparov modules, the Real KK-theory group
$KKR^{m-n}(C_0(X), pt)=KKR(C_0(X), ~^p\IC l^{n,m})$ consist of
equivalence classes of triples $(~^p{\cal H}, ~^p\phi, ~^pF)$
where $^p{\cal H}= ~^p\IC^{\infty}\otimes ~^p\IC l^{n,m}$ is the
Hilbert space over $C_0(pt)\otimes ~^p\IC l^{n,m} \approx \IC
\otimes ~^p\IC l^{n,m}$, $~^p\phi:~^pC_{0}(X) \rightarrow
~^p{\textbf{B}}({~^p{\cal H}})$ is a $\ast$-homomorphism and $^pF$
is a self-adjoint operator in $~^p{\textbf{B}}({~^p{\cal H}})=
~^p{\textbf{B}}(~^p\IC^{\infty})\otimes ~^p\IC l^{n,m}$. On all of
them, the index $p$ means that there is an induced involution
${\cal I}_{9-p}$ (in our case from the orientifold action on the
spacetime) with the properties mentioned above. Also we require
for the tachyon and the scalar fields to be odd and even
respectively under the $\IZ_2$-grading.

In this context, the tachyon is written as
\begin{align}
F =\sum_{A_l\in ~^p\IC l^{n,m}_\text{odd}}T_lA_l,
\end{align}
where $T_l \in ~^p{\textbf{B}}(~^p\IC^{\infty})$ which transforms
on a representation determined by the self-duality condition
$F=F^\dagger$ and the $A_l$ form a basis for $~^p\IC
l^{n,m}_\text{odd}$, which denotes the odd part of the Real
Clifford algebra $~^p\IC l^{n,m}$ (see Appendix C). Similarly, the
unitary transformation $U \in ~^p{\textbf{B}}({~^p{\cal H}})$ on
$^p{\cal H}$, which is a gauge transformation, is even with
respect to the $\IZ_2$-grading determined by $(-1)^{F_L}$. Hence,
such transformation, together with the scalar fields, are written
as
\begin{align}
^p\phi=\sum_{B_l\in ~^p\IC l^{n,m}_\text{even}} \phi_lB_l,
\end{align}
where $ \phi_l \in ~^p{\textbf{B}}(~^p\IC^{\infty})$ and the
corresponding representation is obtained from the condition
$^p\phi=~^p\phi^\dagger$. $B_l$ form a basis for $~^p\IC
l^{n,m}_\text{even}$, which denotes the even part of the Real
Clifford algebra $~^p\IC l^{n,m}.$

Notice as well that the tachyon, scalar fields and the unitary
transformation must be invariant under the orientifold action,
i.e., written in terms of the Clifford algebra elements, they
belong to the so-called fixed point algebra of the corresponding
Clifford Algebra \footnote{It can be shown that an element of the
fixed point algebra of $~^p{\textbf{B}}(~^p\IC^{\infty})$ is,
roughly speaking an infinity real matrix with no involution; then
the condition of belonging to the fixed point algebra of
$~^p{\textbf{B}}({~^p{\cal H}})=
~^p{\textbf{B}}(~^p\IC^{\infty})\otimes ~^p\IC l^{n,m}$ is
equivalent to belong to the fixed point algebra of $~^p\IC
l^{n,m}$ times an infinity real matrix i.e we only need to know
the fixed point algebra of $~^p\IC l^{n,m}$.}. (See Appendix C for
details.)

\subsection{D-branes in orientifolds and Real KK-theory}

With all the necessary ingredients we are in position to construct
the relevant KKR-group which classifies D$d$-brane in the presence
of orientifold planes. As we have seen, one can construct it by
analyzing the D$d$-brane transversal space.

\begin{figure}[t]
\begin{center}
\centering \epsfysize=8cm \leavevmode
\epsfbox{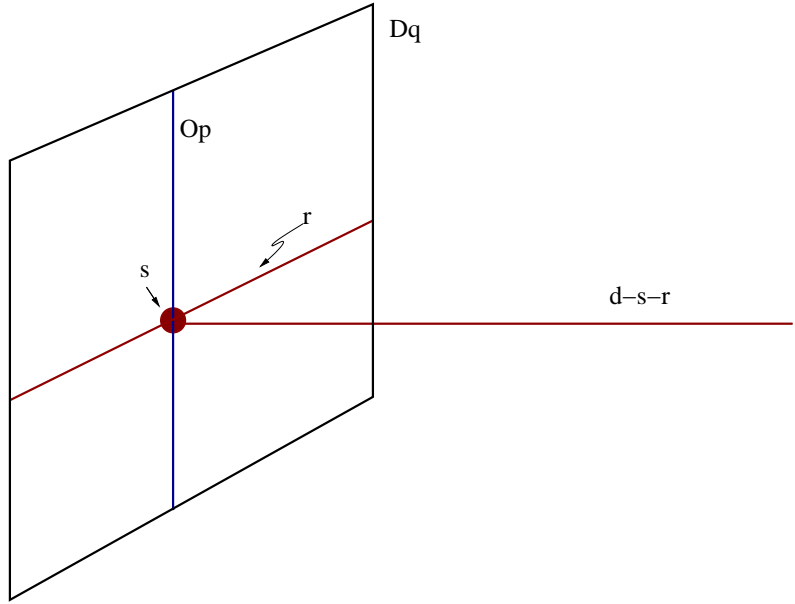}
\end{center}
\caption[]{\small \it A D$d$-brane constructed from tachyon
condensation from unstable D$q$-brane and unstable D-instantons.
The orientifold $Op$ lies inside the D$q$-brane worldvolume.}
\label{Fig:KKR1}
\end{figure}

Let us start by identifying the spaces $X$ and $Y'$. There are two
different configurations according to the relative values between
$q$ and $p$, i.e., whether the plane $Op^-$ is immersed in the
unstable D$q$-brane ($q>p$) or viceversa ($p>q$). Let us start
with the first case as depicted in Fig.\ref{Fig:KKR1}.

We fix our notation by claiming that the final D$d$-brane is
located in coordinates $x^0,\cdots, x^s, x^{p+1}, \cdots ,
x^{p+r}, x^{q+1},\cdots , x^{q+d-s-r}.$ Notice also that our
assumption is that the subspace of the D$d$-worldvolume of
dimension $(s+r)$ is created by the usual tachyon condensation
from the D$q$-brane, while the subspace of dimension $(d-s-r)$ is
gathered from tachyon condensation as in the K-matrix theory.
Therefore, the transversal space $Y'$ to the subspace of dimension
$(r+s)$ is given by $\IR^{(9-q)+(q-p-r), p-s}$, while the subspace
$X$ with dimension $({d-s-r})$ is $\IR^{d-s-r,0}$. Hence it
follows that the KKR-group classifying D$d$-branes in this
configuration is given by
\begin{eqnarray}
KKR(\IR^{d-s-r,0}, \IR^{(9-q)+(q-p-r),
p-s})&=&KKR^{1-q}(\IR^{d-s-r,0},\IR^{q-p-r,p-s}),
 \label{Eq:KKR1}
\end{eqnarray}
where we have used the relations (\ref{isoKKR}) for the last two
terms. We can see that $Y=\IR^{q-p-r,p-s}$

\begin{figure}[t]
\begin{center}
\centering \epsfysize=8cm \leavevmode
\epsfbox{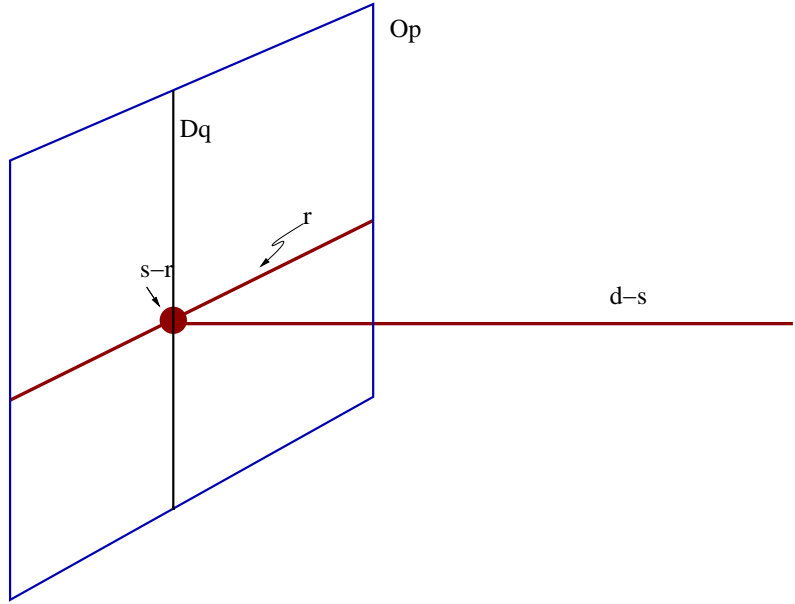}
\end{center}
\caption[]{\small \it D$q$-brane dimensionally lower than the
orientifold plane $Op$.} \label{Fig:KKR2}
\end{figure}

Let us now focus in our second configuration, i.e., the case in
which the D$q$-brane is immersed in the orientifold plane $Op^-$
as depicted in Fig. \ref{Fig:KKR2} ($p>q$). Notice that in this
case, there are some transversal coordinates of the D$d$-brane
with respect to the D$q$-brane which are extended also inside the
orientifold plane. We consider the D$d$-brane to be extended in
coordinates $x^0,x^1,\cdots ,x^{s-r}, x^{q+1},\cdots , x^{q+r},
x^{p+1}, \cdots , x^{p+d-s}$, while the unstable D$q$ and the
orientifold are extended in coordinates labeled by their
dimensions.

Hence the transversal space $Y'$ is $\IR^{9-p,p-q+(q+r-s)}$, while
the space $X$ is given by $\IR^{d-s, r}$ such that the relevant
$KKR$-group is
\begin{align}
KKR(\IR^{d-s, r}, \IR^{9-p, p-q+(q+r-s)})=KKR^{9-2p+q}(\IR^{d-s,r},\IR^{0,q+r-s}),
\label{Eq:KKR2}
\end{align}
where in the last equality we have used the isomorphisms for
$KKR$. Since we are working with $p=1,5,9$, this last group reduces to $KKR^{q-1}(\IR^{d-s,r},\IR^{0,q+r-s})$. Notice that in this way, we can identify the second entrance in the bifunctor $\IR^{0,q+r-s}$ as the D$d$-brane's transversal space within the D$q$-brane.\\

There are actually two special limits we want to consider. In type I theory one has
$p=9$ and we should recover the results given in
\cite{Asakawa:2002nv}. Indeed, in this case ($p>q$) we get that
$s=d$ and from (\ref{Eq:KKR2})
\begin{eqnarray}
KKR^{q-1}(\IR^{d-s,r},\IR^{0,q+r-s})&=&KKR^{q-1}(pt,\IR^{0,q-d})\nonumber\\
&=&KR^{q-1}(\IR^{0,q-d})\nonumber\\
&=&KO^{q-1}(\IR^{q-d})\nonumber\\
&=&KKO^{q-1}(\IR^{d+m-q},\IR^{m}), \label{limit1}
\end{eqnarray}
where $m=q-d+r$ is the codimension between the part of the
D$d$-brane inside of the unstable D$q$-brane system.\\

The second limit we want to check is that of a D$d$-brane located
on top of an orientifold with $p\neq 9$ and for this we can take
the case $q\geq p$. From Fig. 2 this configuration is equivalent
to set $d=s$ and $r=0$ in (\ref{Eq:KKR1}). Hence we have that
$X=pt$ and
\begin{eqnarray}
KKR^{1-q}(\IR^{d-s-r,0},\IR^{q-p-r,p-s})&=&KKR^{1-q}(pt,\IR^{q-p,p-d})\nonumber\\
&=&KR^{1-q}(\IR^{q-p,p-d}), \label{limit2}
\end{eqnarray}
which is in agreement with previous results from section 3.1.

One can ask what kind of extra information (with respect to
K-theory) does we get from these groups. The main point (besides
some more formal statements) is that we can have now a group which
classifies D-branes intersecting orientifold planes. This can be
achieved easily by noticing that Eqs.(\ref{Eq:KKR1}) and
(\ref{Eq:KKR2}) can be written as $KKR(\IR^{d-s,0},\IR^{9-p,
p-s})$ which satisfies
\begin{align}
KKR(\IR^{d-s,0},\IR^{9-p, p-s})= KO(\IR^{2p-2s+d-1}),
\label{reduc}
\end{align}
from which we can see that specific values of $q$ and $r$ are not
important. This means that it does not matter which unstable brane
we select to construct a D$d$-brane, but how many coordinates $s$
of the D$d$-brane are inside the orientifold plane.

\subsubsection{Example 1}
It is easy to check that the orthogonal $KO$-group in
(\ref{reduc}) classifies D-branes in a Type I T-dual version. To
see this, consider for instance a D3-brane in coordinates
($x^0,x^1,x^2,x^3$), and an orientifold $O1^-$ located in
coordinates $x^0, x^1$. By applying T-duality on coordinates
$x^2$-$x^9$, one gets a D7-brane in Type I theory. Such a brane is
classified by $KO(\IR^2)=\IZ_2$. Now let us check if Eq.
(\ref{reduc}) leads us to the same group. In this case, $q\geq p$
and the configuration is similar to that depicted in
Fig.\ref{Fig:KKR1}. It turns out that $s=p=1$ and
\begin{enumerate}
\item $r=2$ and $d-s-r=0$, or \item $r=1$ and $d-s-r=1$, or \item
$r=0$ and $d-s-r=2$.
\end{enumerate}
For all cases, Eq. (\ref{reduc}) gives the group $KO(\IR^2)$ in
agreement with T-duality. The same applies for all different
configuration of D-branes and orientifold planes. The KO-theory
groups from Eq. (\ref{reduc}) classifies the T-dual version in
Type I theory.

One can try to do the same for other type of orientifolds, like the ones with a negative square involution (positive RR charge) and for orientifolds in Type IIA. However, for such cases the related KK groups are not well known from the mathematical point of view. Hence, we can only establish some expected properties for such groups based on physical arguments. We shall comment on these issues in the section \ref{sec:exotic}.\\

\section{Unstable non-BPS D-branes in orientifolds and KKR-theory}
\label{sec:kkrunstable}

We shall follow the criteria in \cite{Asakawa:2002nv} to show that
Eqs. (\ref{Eq:KKR1}) and (\ref{Eq:KKR2}) correctly classify
D$d$-branes in orientifold backgrounds. Hence, we shall extract
the field content of unstable non-BPS D$q$-branes from the
Clifford algebra related to the KKR-group. As we have seen, the
related Clifford algebra is a complex algebra with an antilinear
involution induced by the orientifold action. In this section we
shall obtain the Clifford algebra for each configuration of
non-BPS D$q$-branes and $Op$-planes, and we shall see that the
field content perfectly agrees with that of an unstable non-BPS
brane. Finally we use T-duality to show that the properties of the
non-BPS branes are those expected from non-BPS branes in Type I
theory.

Our proposal for the classification of D$d$-branes in $Op^-$-plane
background is given by Eq. (\ref{Eq:KKR1}) and Eq. (\ref{Eq:KKR2})
according whether $q>p$ or$p>q$.

By using Eq.(\ref{kkrcl}) we have that
$$KKR^{1-q}(X,Y)=KKR_{q-1}(X,Y)$$
\begin{eqnarray}
&=&\left\{\begin{array}{cc}
KKR(C_0(X;\IC )\otimes ~^p\IC l^{0,q-1},C_0(Y;\IC) ) & q-1 > 0 \\
KKR(C_0(X;\IC)\otimes ~^p\IC l^{1-q,0},C_0(Y;\IC)) & 1-q > 0 ,
\end{array}
\right. \label{kkrc}
\end{eqnarray}
for $p<q$ while for $p>q$, we have
$$KKR^{q-1}(X,Y)=KKR_{1-q}(X,Y)$$
\begin{eqnarray}
&=&\left\{\begin{array}{cc}
KKR(C_0(X;\IC),C_0(Y;\IC)\otimes ~^p\IC l^{1-q,0}) & 1-q > 0 \\
KKR(C_0(X;\IC),C_0(Y;\IC)\otimes ~^p\IC l^{0,q-1}) & q-1 > 0 .
\end{array}
\right. \label{kkrc2}
\end{eqnarray}

We shall use the above formulae as definition of the KKR-theory
groups for D-brane classification. This choice is taken in order to recover the
convention in \cite{Asakawa:2002nv} where $p=9$ and consequently $q \leq p$. Since in this case the involution is trivial,
(\ref{kkrc2}) reduces to the definition for $KKO^{-n}(X,Y)$ used
in \cite{Asakawa:2002nv}. In that sense, we classify non-BPS
D$d$-branes by making them to coincide with the unstable
D$q$-system, implying that $X=pt$ and that $Y=pt$. By this
assumption we can safely conclude that all information about these
non-BPS D-branes relies on the corresponding Clifford algebras.
All what we need to specify is the involution action on the
Clifford algebra generators.

\subsection{The Real involution and orientifolds}
Let us describe explicity how  the real involution acts on the
Clifford generators induced by the orientifold $Op^-$-plane.

The complexified Real Clifford algebra is defined as
\begin{align}
^p\IC l^{n,m}=~^p \left(Cl^{n,m}\otimes \IC\right).
\end{align}
Since the involution acts as conjugation on the complex part we
can write (see Appendix C for details)
\begin{align}
^p  \left(Cl^{n,m}\otimes \IC\right)= ~^p Cl^{n,m} \otimes
\overline{\IC},
\end{align}

\noindent where $\overline{\IC}$ denotes the field of complex
numbers with Real involution defined by usual complex conjugation
and $~^pCl^{n,m}$ denotes the Clifford algebra $Cl^{n,m}$ with
some Real involution (again, this involution is determined by the
orientifold plane on the generators of the algebra and extended by
linearity).Thus, it suffices to study the involution in the real
part $^pCl^{n,m}$. Hence, we shall concentrate on how to fix the
involution inhereted from the orientifold $Op^-$-plane on the
generators of the real Clifford algebra.

According to Eqs. (\ref{kkrc}) and (\ref{kkrc2}) the complex
Clifford algebras with involution we use, are of the form $^p \IC
l^{n,0}$ or $^p\IC l^{0,n}$. Hence we shall concentrate on the
involution on their associated orthogonal (real) Clifford
algebras, whose generators can be  identified with spatial
coordinates via the vector space isomorphism
\begin{align}
Cl^{n,0} \cong \Lambda^{\ast} \mathbb{R}^{n} \cong Cl^{n,0}.
\end{align}
Let us consider the case in which $1-q<0$ such that the related
Clifford algebra is $^p \IC l^{0, q-1}$. By the above isomorphism
we identify the generators  $e_i$, ($i=1,\cdots , q-1$) of the
Clifford algebra $Cl^{0,q-1}$  with vectors of the little group
$SO(q-1)$ of a D$q$-brane.

Hence, the involution inhereted form the orientifold $p$-plane,
denoted as ${\cal I}_{9-p}$ acts on the generators of the complex
Clifford algebra as in the longitudinal coordinates $x^i$ to the
D$q$-brane. Because of this, the involution depends on the
relative value between $p$ and $q$. Then, if $q<p$ we consider a
D$q$-brane inside the orientifold plane, as in the following
configuration
\begin{center}
\begin{tabular}{c|cccccccccccccc}
&0&1&2&$\cdots$&$q$-1&$q$&$q$+1&$\cdots$&$p$-1&$p$&$p$+1&$\cdots$&8&9\\
\hline\\
$Op^-$&$-$&$-$&$-$&$-$&$-$&$-$&$-$&$-$&$-$&$-$&$\times$&$\times$&$\times$&$\times$\\
$Dq$&$-$&$-$&$-$&$-$&$-$&$-$&$\times$&$\times$&$\times$&$\times$&$\times$&$\times$&$\times$&$\times$\\
\end{tabular}
\end{center}
inducing a trivial involution on all the Clifford algebra
generators
\begin{align}
{\cal I}_{9-p}(e_i)= e_i \quad \text{for all $i$}.
\end{align}
On the other hand, if $q>p$, the unstable D$q$-brane is located as

\begin{center}
\begin{tabular}{c|cccccccccccccc}
&0&1&2&$\cdots$&$p$-1&$p$&$p$+1&$\cdots$&$q$-1&$q$&$q$+1&$\cdots$&8&9\\
\hline\\
$Op^-$&$-$&$-$&$-$&$-$&$-$&$-$&$\times$&$\times$&$\times$&$\times$&$\times$&$\times$&$\times$&$\times$\\
$Dq$&$-$&$-$&$-$&$-$&$-$&$-$&$-$&$-$&$-$&$-$&$\times$&$\times$&$\times$&$\times$\\
\end{tabular}
\end{center}
and the involution is given by
\begin{equation}
\qquad \qquad \qquad \qquad {\cal I}_{9-p}(e_i)=\left\{
\begin{array}{cc}
e_i \quad \text{for} \quad i=1,...,p-1,\\
-e_i \quad \text{for} \quad i=p,...,q+1.
\end{array}
\right.
\end{equation}

By Bott periodicity and Eqs.(\ref{kkrc}) and (\ref{kkrc2})  we have
\begin{align}
KKR(X, ~^p \IC l^{0,q-1})=KKR(X, ~^p\IC l^{9-q, 0}).
\end{align}
 In this way, we can use instead the $(9-q)$ generators $e_i$ of $^p \IC l^{9-q,
0}$, with $i=q+1, \cdots , 9$ which are identified with the
transversal coordinates to the D$q$-brane. The involution is again
dependent on the relative values between $q$ and $p$. For $q<p$
(see Fig. \ref{Fig:KKR2}) we have
\begin{equation}
\qquad \qquad \qquad {\cal I}_{9-p}(e_i)=\left\{
\begin{array}{cc}
e_i \quad \text{for} \quad i=1,\cdots ,p-q,\\
-e_i \quad \text{for} \quad i=p-q+1,...,9-q,
\end{array}
\right. \label{inv}
\end{equation}
while for $q>p$ we have
\begin{align}
{\cal I}_{9-p}(e_i)= -e_i \qquad \text{for all $i$}.
\end{align}

Therefore, one sees that for $q>2$, we have at least two different
ways to identity the Clifford algebra generators with spatial
coordinates i.e. internal or transversal coordinates to the
D$q$-brane system. For each identification there are two choices
for the involution on the Clifford generators, depending on the
relative value of $q$ and $p$. However we also see that for $q<p$
is simpler to establish the identification with internal
coordinates to the D$q$-brane, while for the case $p<q$ is the
opposite. We shall adopt this identification henceforth.

Although the identifications are not so geometric for $q<2$, we
have similar involutions. For $q=-1$ the relevant Clifford algebra
is $^p Cl^{2,0}$ and the involution acts on the generators as
${\cal I}_{9-p}(e_i)= e_i$ ($i=1,2$). Similarly for $q=0$, the
Clifford algebra is $^p Cl^{1,0}$ and the involution acts also
trivially on the generator.

\subsection{Non-BPS D-branes in orientifold backgrounds}
Now, we are going to get the representations of the tachyon, gauge
and scalar fields from the corresponding Clifford algebras,
following the procedure used in \cite{Asakawa:2002nv}, and we will
show that they correspond to the properties of unstable non-BPS
D$d$-branes classified by the groups in Eqs. (\ref{Eq:KKR1}) and
(\ref{Eq:KKR2}). Due to Bott periodicity in $KKR^n(X,Y)\sim
KKR^{n\pm 8}(X,Y)$, all cases are considered within the range
$-4\leq n \leq 4$. However, in contrast with D-branes in Type I
theory, the involution acts different for a D$q$-brane than for a
D$(q+8)$-brane. Notice as well that, although Eqs. (\ref{Eq:KKR1})
and (\ref{Eq:KKR2}) do not depend on $p$ (in these kinds of
non-BPS branes), the involution does.

\subsubsection{Example 1}
Consider for instance the case of a non-BPS D8-brane and an
$O1^-$-plane in a configuration as follows
\begin{center}
\begin{tabular}{c|cccccccccc}
&0&1&2&3&4&5&6&7&8&9\\
\hline\\
$O1^-$&$-$&$-$&$\times$&$\times$&$\times$&$\times$&$\times$&$\times$&$\times$&$\times$\\
$D8$&$-$&$-$&$-$&$-$&$-$&$-$&$-$&$-$&$-$&$\times$.\\
\end{tabular}
\end{center}
The corresponding group is $KKR^{-7}(pt, pt)\sim KKR^{1}(pt,pt)$
with an associated Real Clifford algebra $^1\IC l^{1,0}$. The
action of the involution on the generator of $^1Cl^{1,0}$ is given
by
\begin{align}
{\cal I}_8(e_1)=-e_1.
\end{align}
This determines the fixed point algebra for $^1\IC l ^{1,0}$ and
hence, the corresponding representation for the tachyon, gauge and
scalar fields. By imposing the condition ${\cal I}_8(a)=a$ for
$a\in ~^1\IC l^{1,0}$, one gets that
\begin{align}
\left(~^1\IC l^{1,0}\right)_\text{fix}= Cl^{0,1},
\end{align}
which fixes the tachyon $T$ and the scalar field $\phi$ to be
symmetric tensor representations ${\Ysymm}$ of the gauge group
$O(\infty)$. As it was shown in \cite{Asakawa:2002nv}, these
results correspond to the field content of an unstable non-BPS
D2-brane in Type I theory. This is in agreement with formula
(\ref{Eq:KKR1}) since for this case\footnote{Actually, as we shall
see, similar conditions hold for all unstable non-BPS D-branes.}
$p=s=1$, $d=q=8$ and $r=7$, and the relevant KKR group is given by
\begin{align}
KKR^{-7}(pt,pt)= KO(\IR^7)= 0,
\end{align}
which indeed is the K-theory group which classifies D2-branes in
Type I theory. One can as well check that under T-duality on
transversal coordinates to the $O1^-$-plane, the unstable D8-brane
transforms into a D2-brane in Type I theory. Notice that the
involution does not change for $p=5$, for which we get the same
field content for a D8 in an $O5^-$-plane.

\subsubsection{Example 2}
Contrary to the case in Type I theory, the field content for a
non-BPS D0-brane in an $O1^-$-plane
\begin{center}
\begin{tabular}{c|cccccccccc}
&0&1&2&3&4&5&6&7&8&9\\
\hline\\
$O1^-$&$-$&$-$&$\times$&$\times$&$\times$&$\times$&$\times$&$\times$&$\times$&$\times$\\
$D0$&$-$&$\times$&$\times$&$\times$&$\times$&$\times$&$\times$&$\times$&$\times$&$\times$\\
\end{tabular}
\end{center}
should not be the same than for a D$8$. This is obtained by
realizing that for a D0-brane, although the Real Clifford algebra
also is $^1\IC l^{1,0}$, the involution on the single one
generator $e_1$ is trivial, ${\cal I}_8(e_1)=e_1$. This implies
that
\begin{align}
\left(~^1\IC l^{1,0}\right)_\text{fix}= Cl^{1,0}.
\end{align}
Therefore, the tachyon field $T$ and the scalar field $\phi$ are
antisymmetric ${\Yasymm}$ and symmetric ${\Ysymm}$ tensor
representations, respectively, of the gauge group $O(\infty)$
\cite{Asakawa:2002nv}. This field content is precisely that of an
unstable non-BPS D0-brane in Type I theory. This also is in
agreement with formula (\ref{Eq:KKR2}) in which $r=s=q=d=0$ and
$p=1$, implying
\begin{align}
KKR^{-1}(pt,pt)=KO(\IR^1)= \IZ_2,
\end{align}
which classifies D8-branes in Type I theory. Indeed, the
configuration of a D0-brane in an $O1^-$-plane is T-dual to a D8
in an $O9^-$-plane. For $p=5$, the involution is the same and we
get the same group.


\subsubsection{Example 3}
Another interesting situation presents for $q=5$, i.e., D5-branes
in $O1^-$ and $O5^-$-planes. The Real Clifford algebra is given by
$^p\IC l^{4,0}$ for $p=1,5$. In this case, the involution acts as
${\cal I}_4(e_i)=-e_i$ for $i=2,3,4,5$. As a consequence, the
fixed point algebra is $Cl^{0,4}$. For this case, we can also take
the Real Clifford algebra as $^p\IC l^{4,0}=~^p\IC l^{0,4}$.
However the involution acts trivially on the corresponding
generators. The fixed point algebra is then $Cl^{4,0}$. It is easy
to check that $Cl^{4,0}=Cl^{0,4}$. Hence, as it was shown in
\cite{Asakawa:2002nv}, the tachyon and scalars fields transforms
in the bifundamental and antisymmetric tensor representations of
the gauge group $Sp(\infty)\times Sp(\infty)$. This is the field
content of a pair D$5$-$\overline{\hbox{D}5}$ branes in Type I
theory, which agrees with the result given by
\begin{align}
KKR^{-4}(pt,pt)=KO(\IR^4)=KSp(pt)=\IZ.
\end{align}
The complete set of Real Clifford algebras for all unstable
non-BPS branes is summarized in Table \ref{Tabla:todo}. The
representations and gauge groups for each case are recovered from
the results shown in \cite{Asakawa:2002nv} just by computing the
fixed point algebras, as in the previous examples. For completness
we summarize such results in Appendix C(see Table
\ref{Tabla:japon}).

\begin{table}
\begin{center}
\begin{tabular}{|c|c||c|c|c|c|c|}
\hline
&&&&&&\\
&D$d$&$^p\IC l^{n,m}$&$(^pCl^n)_\text{fix}$&$KKR^n$&$KO^n(pt)$&T-dual in Type I\\
\hline\hline
&&&&&&\\
$p=1$&D(-1)&$^1\IC l^{2,0}$&$Cl^{2,0}$&$KKR^{-2}$&$KO^{-2}=\IZ_{2}$&D7\\
$p=5$&     &$^5\IC l^{2,0}$&          &$KKR^{-10}$&$KO^{-10}=\IZ_{2}$&D(-1)\\
\hline
&&&&&&\\
&D0&$^1\IC l^{1,0}$&$Cl^{1,0}$&$KKR^{-1}$&$KO^{-1}=\IZ_{2}$&D8\\
&  &$^5\IC l^{1,0}$&          &$KKR^{-9}$&$KO^{-9}=\IZ_{2}$&D0\\
\hline
&&&&&&\\
&D1-$\overline{\hbox{D}1}$&$^1\IC l^{1,1}$&$Cl^{1,1}$&$KKR^{0}$&$KO^{0}=\IZ$&D9-$\overline{\hbox{D}9}$\\
&  &$^5\IC l^{1,1}$&          &$KKR^{-8}$&$KO^{-8}=\IZ$&D1-$\overline{\hbox{D}1}$\\
\hline
&&&&&&\\
&D2&$^1\IC l^{0,1}$&$Cl^{1,0}$&$KKR^{-1}$&$KO^{-1}=\IZ_{2}$&D8\\
&  &$^5\IC l^{0,1}$&$Cl^{0,1}$&$KKR^{-7}$&$KO^{-7}=0$&D2\\
\hline
&&&&&&\\
&D3&$^1\IC l^{0,2}$&$Cl^{2,0}$&$KKR^{-2}$&$KO^{-2}=\IZ_{2}$&D7\\
&  &$^5\IC l^{0,2}$&$Cl^{0,2}$&$KKR^{-6}$&$KO^{-6}=0$&D3\\
\hline
&&&&&&\\
&D4&$^1\IC l^{0,3}$&$Cl^{3,0}$&$KKR^{-3}$&$KO^{-3}=0$&D6\\
&  &$^5\IC l^{0,3}$&$Cl^{0,3}$&$KKR^{-5}$&$KO^{-5}=0$&D4\\
\hline
&&&&&&\\
&D5-$\overline{\hbox{D}5}$&$^1\IC l^{0,4}$&$Cl^{4,0}$&$KKR^{-4}$&$KO^{-4}=\IZ$&D5+$\overline{\hbox{D}5}$\\
&           &$^5\IC l^{0,4}$&          &          &         &\\
\hline
&&&&&&\\
&D6&$^1\IC l^{3,0}$&$Cl^{0,3}$&$KKR^{-5}$&$KO^{-5}=0$&D4\\
&  &$^5\IC l^{3,0}$&          &          &         &\\
\hline
&&&&&&\\
&D7&$^1\IC l^{2,0}$&$Cl^{0,2}$&$KKR^{-6}$&$KO^{-6}=0$&D3\\
&  &$^5\IC l^{2,0}$&          &          &         &\\
\hline
&&&&&&\\
&D8&$^1\IC l^{1,0}$&$Cl^{0,1}$&$KKR^{-7}$&$KO^{-7}=0$&D2\\
&  &$^5\IC l^{1,0}$&          &          &         &\\
\hline
\end{tabular}
\caption{$KKR$-groups and their related Clifford algebras, fixed
point algebras and $KO$-theory groups for unstable D$q$-branes in
$O1^-$ and $O5^-$-planes. The empty entries stand for the same
expressions as the preceding row.} \label{Tabla:todo}
\end{center}
\end{table}


\subsection{D$q$-branes from D-instantons in orientifold backgrounds}
As we have said, we shall follow the criteria in
\cite{Asakawa:2002nv} to test the validity of formulae
(\ref{Eq:KKR1}), (\ref{Eq:KKR2}). For that we are going to show
explicitly the construction of a D$d$-brane from an infinitely
many number of instantons in the presence of an orientifold plane
$O1^-$ or $O5^-$. In \cite{Asakawa:2002nv} the authors found that
the tension of D$d$-branes in Type I theory are related to the
size(dimension of the representation) of $SO(d)$ gamma matrices.
In the case of lower dimensional orientifold planes, we shall get
a similar relation.

The strategy in \cite{Asakawa:2002nv} adapted to our case is as
follows. An explicit configuration representing a D$d$-brane is
gathered by constructing the corresponding configuration in Type
IIB, based on D-instanton-anti-D-instanton, which survives after
the orientifold projection.

Hence, since the relevant Real Clifford algebra related to a
system of D$(-1)$-$\overline{\hbox{D}(-1)}$ is $\IC l^{2,0}$, and
being the tachyon field odd with respect to the $\IZ_2$-grading,
it can be written as
\begin{align}
F=T_1\widehat{e}_1+T_2\widehat{e}_2,
\end{align}
where $\widehat{e}_1, \widehat{e}_2 \in \IC
l^{2,0}_{\text{odd}}=(Cl^{2,0}_\text{odd}\otimes\IC)$, and $T_1$
and $T_2$ are real fields. Besides this, the tachyon field is
self-dual ($F=F^\dag$) and is invariant under the involution
${\cal I}_{9-p}$, i.e.
\begin{align}
{\cal I}_{9-p}(F)=F,
\end{align}
which makes it belongs to the fixed point algebra of the
corresponding Real Clifford algebra. Then, the tachyon field can
also be written as
\begin{align}
F=T_ae_1+T_be_2, \label{taqab}
\end{align}
where $T_a$ and $T_b$ are complex fields and $e_1, e_2 \in
Cl^{2,0}_\text{odd}$. Defining the field $T= T_a+T_be_1\wedge e_2$
one gets that $T=-T^\dag$ due to the self-duality condition on
$F$. In particular, we observe that for an $O9^-$-plane, the
involution acts trivially on all Clifford algebra generators. This
implies that
$\rm{Im}~\textit{T}_{\textit{a}}=\rm{Im}~\textit{T}_{\textit{b}}=0$
and that $T=-T^\text{T}$.

Now, since for a D$d$-brane constructed from instantons, the
tachyon field also reads
 \begin{align}
 F=\mu\sum_{i=0}^d p_i\otimes \Gamma^i,
 \end{align}
comparing with Eq. (\ref{taqab}) we conclude that
\begin{align}
T_a=\partial_0\otimes\gamma^0, \nonumber\\
T_b=\partial_j\otimes\gamma^j_d,
\end{align}
with $(\gamma^\mu_{d+1})^\dag=\gamma^\mu_{d+1}$ being hermitian
$\gamma$-matrices, which in the abscence of an orientifold plane,
are irreducible hermitian $SO(d+1)$ gamma matrices. In the
presence of orientifold planes $O5^-$ and $O1^-$, it turns out
that the involutions ${\cal I}_4$ and ${\cal I}_8$ act trivially
on the generators $e_1$ and $e_2$ (as in the Type I case). This
renders the gamma-matrices to split into $\gamma^0=I$ and $SO(d)$
gamma matrices $\gamma^i$ ($i=1,\dots ,d$) with the latter forming
a real representation of $Cl^{0,d}$. Using this information we can
compare the size of the tachyon in Type IIB and in the presence of
orientifold planes. The ratio does not depend on $p$, implying
that the tension (and size) of a D$d$-brane in an $O5^-$, $O1^-$
and $O9^-$ (as in the configurations considered in the previous
section) is twice than that in Type IIB for $d=3,4,5,6,7$. Notice
that for $p=1,5$ the D$d$-branes with twice the tension than in
Type IIB are T-duals to those in Type I theory which also have
twice the tension as their counterparts in Type IIB. This is shown
in Table 2. Notice as well, as it was pointed out in
\cite{Asakawa:2002nv}, that this is consistent with the
construction of D-branes in Type I theory, since those branes in
an $Op^-$-plane with twice the tension than in Type IIB, are
T-duals to Type I D-branes constructed from two Type IIB branes or
a pair of brane-antibrane.

\begin{table}
\begin{center}
\begin{tabular}{|c||c|c|c|c|c|c|c|c|c|c|c|}
\hline
D$d$&D0&D1&D2&D3&D4&D5&D6&D7&D8&D9\\
\hline
Size in IIB&1&1&2&2&4&4&8&8&16&16\\
Size in IIB + O$p$-plane, $p=1,5$&1&1&2&4&8&8&16&16&16&16\\
T-dual into Type I ($p=1$)&D8&D7&D8&D7&D6&D5&D4&D3&D2&D1\\
T-dual into Type I ($p=5$)&D0&D1&D2&D3&D4&D5&D4&D3&D2&D1\\
\hline
\end{tabular}
\caption{Relative dimension of the representation between gamma
matrices related to D$d$-branes in Type IIB and in
$Op^-$-backgrounds with $p=1,5$.} \label{Tabla:size}
\end{center}
\end{table}

This is our last test to show that indeed, KKR-theory truly
classifies D-branes charges in (the provided) orientifold
backgrounds.

\section{A proposal for clasification of D-branes in $Op^+$-planes}
\label{sec:exotic}

In \cite{Asakawa:2001vm} and \cite{Asakawa:2002nv} KK-theory and
KKO-theory are used to classify D-branes in Type II and Type I
superstring theories respectively. Also in this paper we have
extended this classification to orientifold backgrounds in Type
IIB string theory by using KKR-theory.

Then, it is natural to think about the possibility of other
KK-theories \footnote{At least those KK-theories related with
K-theories classifying consistent stringy backgrounds.}, extending
the K-theory classification of superstring theories in different
backgrounds than those appearing in this paper. In particular, we
focus on Type IIB O$p^{+}$ orientifolds (the involution induced on
the Chan-Paton bundles is $\tau^{2}=-1$) which are classified by
quaternionic K-theory, denoted KH \cite{Gukov:1999yn} and
symplectic $USp(32)$ (IIB + O9$^{+}$) string theory proposed in
\cite{Sugimoto:1999tx} which is classified by symplectic K-theory,
denoted KSp.

We focus on these particular backgrounds because their associated
K-theories have close relation with KO and KR theories\footnote{In
\cite{Asakawa:2002nv}, though they do not make explicit mention of
KKSp theory, they use the relation between KO and KSp theories to
conclude that $USp(32)$ theory is classified by $KKO^{q+3}$, where
$q$ is de dimension of the unstable $D$-brane in the $USp(32)$
theory. But $USp(32)$ string theory is a consistent theory, then
there should exist KKSp theory, which should be related to
KKO-theory in a suitable way to achieve the KKO-groups proposed in
\cite{Asakawa:2002nv}.} and consequently, we can conjecture some
relations that their corresponding KKH and KKSp theories must
satisfy.

For this purpose, we first write some properties and relations
between KH, KSp and KO theories:

\begin{equation}
 \qquad \qquad \qquad \qquad \qquad \qquad KH(X) \simeq
 KH^{-8}(X),
\label{kh1}
\end{equation}

\begin{equation}
\qquad \qquad \qquad \qquad KH^{p,q}(X) \simeq KH^{p+1,q+1}(X)
\simeq KH^{p-q}(X), \label{kh2}
\end{equation}

\begin{equation}
 \qquad \qquad \qquad \qquad \qquad \qquad KH(X_{R}) \simeq
 KSp(X_{R}),
\label{kh3}
\end{equation}

\begin{equation}
\qquad \qquad \qquad \qquad \qquad \qquad KSp(S^{n}) \simeq
KO(S^{n+4}). \label{kh4}
\end{equation}

In (\ref{kh3}), $X_{R}$ is the fixed point set of the involution
of the spacetime and this property reflects the fact that
KH-theories are T-duals of $USp(32)$ theory, with the involution
acting on the dualized coordinates. For example if we start with
$USp(32)$ theory and we do not make any T-duality, then the
involution does not act at all in the spacetime; so in this case
the fixed spacetime is the fixed point set of the involution; in
this way $KSp(X)=KH(X)$ and $USp(32)$ theory can be regarded as
(IIB + O9$^{+}$)-string theory, in the same way Type I string
theory can be seen as (IIB + O9$^{-}$)-string theory.

The most important property we shall assume in all KK-theory
groups $KK^{-n}(X,Y)$ proposed here is that when either $X$ or $Y$
is the one point space, they reduce to the respective K-theory and
K-(analytic) homology functors\footnote{As in the case of
KR-homology, there should be a suitable definition of topological
K-homology and it must be possible to prove the equivalence with
the analytical K-homology defined above.}. One consequence of this
property is that our KK-functors must preserve the original
periodicity of their K-functors, i.e $n$ $\mathit{mod}$ $8$
periodicity.

Let us start with (IIB + O9$^{+}$) backgrounds, i.e. KKH-theory.
Both the crucial formula (\ref{isoKKR}) of KKR-theory and the
similar property (\ref{kh2}) of KH-theory shared by KR-theory
allowed us to compute the KKR-groups and to confirm our proposal;
then we also assume that KKH-theory should obey a similar
property:

\begin{equation}
\qquad $$KKH^{k}(X,Y)$$=$$KKH^{k+p-q}(X\times
\mathbb{R}^{p,q},Y)$$=$$KKH^{k-p+q}(X,Y \times
\mathbb{R}^{p,q}).$$ \label{kkhbp}
\end{equation}

Suppose we have a configuration similar to that of Fig.~2 (for our
present purposes it is enough to restrict our attention to this
system; but it is straightforward to adapt the following arguments
for the configuration of Fig.~3. In analogy with the O$p^{-}$
orientifolds, we propose that the KKH-group classifying stable
D-brane configurations is given by

\begin{equation}
 \qquad \qquad  \qquad \qquad \quad KKH(\mathbb{R}^{d-s-r, 0},
  \mathbb{R}^{(9-q)+(q-p-r),p-s}).
\label{pos3}
\end{equation}

In this way the calculations are identical to the ones that lead
to (\ref{Eq:KKR1}); so we have

\begin{equation}
   \qquad KKH(\mathbb{R}^{d-s-r, 0},
  \mathbb{R}^{(9-q)+(q-p-r),p-s})=KKH^{1-q}(\mathbb{R}^{d-s-r, 0},
  \mathbb{R}^{q-p-r,p-s}),
\end{equation}
which can be written, by using (\ref{kh1})-(\ref{kh4}), in the
following way:

\begin{equation}
KKH^{1-q}(\mathbb{R}^{d-s-r, 0},
  \mathbb{R}^{q-p-r,p-s})=KSp(\mathbb{R}^{2p-2s+d-1})=KO(\mathbb{R}^{2p-2s+d+3}).
\label{kkhkspko}
\end{equation}

If we take $r=0$ and $d=s$, then the stable D$d$-brane is located
on top of the orientifold plane and (\ref{kkhkspko}) reduces to:

\begin{equation}
\qquad KKH^{1-q}(\mathbb{R}^{0, 0},
  \mathbb{R}^{q-p,p-d})=KSp(S^{2p-d-1})=KO(S^{2p-d+3}),
\label{kh5}
\end{equation}

\noindent which is precisely Gukov's prescription for D-branes
located on top of O$p^{+}$ orientifolds. Then the basic properties
of KKH-groups mentioned above are enough to carry on the
classification of stable D-branes in Op$^{+}$ orientifolds.

To construct the corresponding $\textit{``quaternionic Kasparov
module''}$ the first step is to define a \textit{``quaternionic
C$^{*}$-algebra''}; which means a Banach $\ast$-algebra $A$ over
the quaternionic field such that, the $C^{*}$-equation $\| x^{*}x
\| = \|x\|^{2}$ holds for any $x \in A$.

Then, one can follows the path traced in \cite{kaspa:1981wl} by
substituting the fields $\mathbb{R}$ or $\mathbb{C}$ by
$\mathbb{H}$; and the complex, real and Real Clifford algebras by
the quaternionic Clifford algebras $\IC l^{n,m}_{H}$
\footnote{$\IC l^{n,m}_{H}$ is defined as the tensor product of
the real Clifford Algebra $Cl^{m,n}$ with the quaternionic field,
i.e $\IC l^{n,m}_{H} = Cl^{m,n} \otimes \mathbb{H}$.} endowed with
some $C^{*}$-algebra structure. Of course, along the way there may
be some subtleties associated with the specific properties of
$\mathbb{H}$, such as noncommutativity.

Now, we turn to $USp(32)$ string theory. Suppose that in this
theory we have a configuration similar to the one described in the
paragraph above equation (\ref{kk1}). Then we postulate (in
accordance with (\ref{kk2}))that stable D-branes are classified by

\begin{equation}
\qquad \qquad \qquad \qquad \qquad
{KKSp}^{q-1}(\mathbb{R}^{d+s-q}, \mathbb{R}^{s}).
 \label{ksp1}
\end{equation}
In order for (\ref{ksp1}) to reproduce the $K$-theory group of the
transverse space of the $Dd$-brane, we postulate the following
property analogous to (\ref{kkbp}):

\begin{equation}
\qquad $$KKSp^{k}(X,Y)$$=$$KKSp^{k-n}(X\times
\mathbb{R}^{n},Y)$$=$$KKSp^{k+m}(X,Y \times \mathbb{R}^{m}).$$
\label{kkSpbp}
\end{equation}

\noindent In this way we get

 $$KKSp^{q-1}(\mathbb{R}^{d+s-q}, \mathbb{R}^{s})= KSp(
 \mathbb{R}^{9-p})$$

 \begin{equation}
\qquad \qquad \qquad \quad
=KO^{-4}(\mathbb{R}^{9-p})=KKO^{q+3}(\mathbb{R}^{d+s-q},
 \mathbb{R}^{s}).
\label{ksp2}
 \end{equation}

From the above equation we reproduce the claim in
\cite{Asakawa:2002nv} that $D$-branes in USp(32) string theory are
classified by $KKO^{q+3}(X,Y).$ So, we claim that

\begin{equation}
\qquad \qquad \qquad KKSp^{i}(X,Y)=KKO^{i+4}(X,Y)=KKO^{i-4}(X,Y).
 \label{ksp3}
\end{equation}

\subsection{An application: exotic orientifolds}
We know that for $p<6$ there are a variatey of orientifold planes,
characterized by their RR and NS-NS charge \cite{Witten:1998xy,
Hanany:2000fq, Bergman:2001rp}. It is interesting to realize that
a cohomological classification of the RR and NS fluxes, tells us
that there are at least 4 different types of orientifold planes
for $p<6$ but only 3 in a K-theoretical classification
\cite{Moore:1999gb, Bergman:2001rp}\footnote{Actually, if one
consider an S-dual version of the conecction between cohomology
and K-theory (called the Atiyah-Hirzebruch Spectral Sequence)
there are just two different types of orientifolds classified by
K-theory \cite{GarciaCompean:2002kc, LoaizaBrito:2003gz}.}.

\begin{figure}[t]
\begin{center}
\centering \epsfysize=8cm \leavevmode
\epsfbox{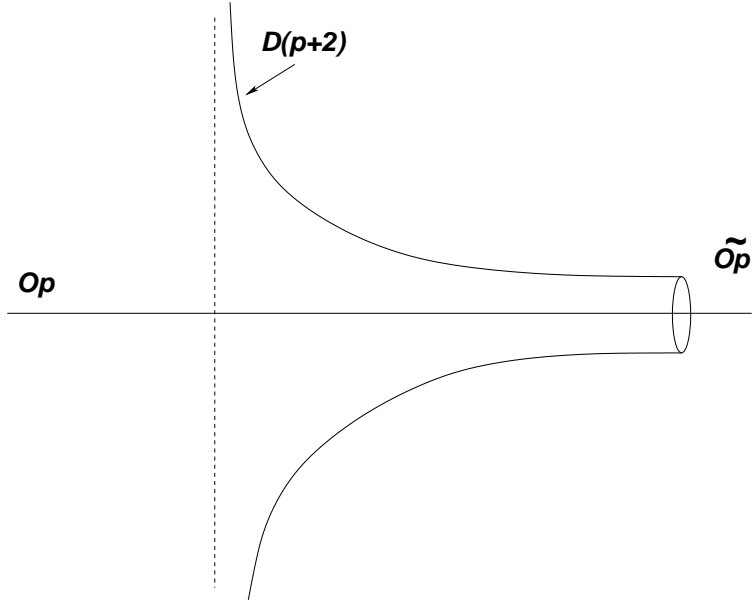}
\end{center}
\caption[]{\small \it Brane realization of
$\widetilde{Op}$-planes.} \label{exotic}
\end{figure}

At the level of cohomology, there are two different types of
orientifold related to RR fluxes. They are classified by the
torsion part of the group $H^{6-p}(\IRP^{8-p},\ItZ)=\IZ_2$ which
is interpreted as a half-shift in RR charge,defining the exotic
orientifold planes $\widetilde{Op}$. The brane realization of this
type of orientifold plane $Op$ is depicted in figure \ref{exotic},
where roughly speaking, an exotic $\widetilde{Op}$-plane is
constructed by wrapping a D$(p+2)$-brane in a two-cycle of the
transverse space of $\IRP^{8-p}$ an $Op$-plane .

However, it can be shown \cite{Bergman:2001rp} that a
K-theoretical classification of RR fields gives more information
such as an explanation for the relative charge between different
types of orientifold planes. In this context $Op^-$ and
$Op^+$-planes are classified (through their RR fields) by
$KR^{p-10}(\IS^{9-p,0})$ and
$KR^{p-6}(\IS^{9-p,0})=KH^{p-10}(\IS^{9-p})$ respectively. For
$p=1,5$ we have the values
\begin{eqnarray}
\qquad \qquad \qquad \qquad O1^-:\qquad KR^{-1}(\IR^{8,0})&=&\IZ\oplus\IZ_2\, \nonumber\\
O1^+:\qquad KR^{-5}(\IR^{8,0})&=&\IZ, \nonumber\\
O5^-:\qquad KR^{-5}(\IR^{4,0})&=&\IZ, \nonumber\\
O5^+:\qquad KR^{-1}(\IR^{4,0})&=&\IZ\oplus\IZ_2. \label{Kop}
\end{eqnarray}
The abscence of a torsional part in the group for $O5^-$-planes,
is interpreted (via the Atiyah Hirzebruch Spectral Sequence) as a
shift in the RR charge by a half-unit, explaining the relative
fractional charge between them and the exotic ones denoted
$\widetilde{O5^{-}}$. In this sense is easy to see that
$\widetilde{O5^+}$ has the same RR charge than $O5^+$. For the
case of $O1$-planes, we have exactly the same situation, although
there are extra $\IZ_2$ constributions from cohomology. Once we
compare this information with the corresponding K-theory groups,
we arrive to the same conclusions as for the $O5$-planes
\cite{Bergman:2001rp}.

Hence, although the existence of these exotic orientifolds comes from cohomology and a more accurate description about their RR charges is given by K-theory, this is actually a classification of RR fields. The D-brane realization of exotic orientifolds suggests on the other hand, a K-theory classification of D-brane (charges). Since an $\widetilde{Op}$-plane is constructed by a D$(p+2)$-brane wrapping a two-cycle transversal to the orientifold, it should be enough to classify such configuration of branes in an orientifold background to elucidate their existence. This is precisely what KKR-theory does at least for $p=1,5,9$.

Consider for instance the case of an $O5^-$-plane and a D7-brane wrapping a two-cycle transversal to the orientifold plane. Let us take the configuration given by $q<p$ (the same result can be obtained by taking $p<q$). Hence we have $d-s=2$ and $s=5$. The related KKR-theory group is
\begin{align}
KKR^{-6}(pt,pt)=KO(\IR^6)=0,
\end{align}
which tells us that there is no extra contribution in K-theory to the $O5^-$-planes. On the other hand, for the $O1$-planes, we have that $d-s=2$ and $s=1$. The related KKR-theory group is then
\begin{align}
KKR^{-2}(pt,pt)=KO(\IR^2)=\IZ_2,
\end{align}
which is also in agreement with (\ref{Kop}). Finally, we can check
that for the cases of $O5^+$ and $O1^+$, the proposed KKH-theory
groups gives the expected results. For the $O5^+$-plane we have
that the relevant KK-theory group is
\begin{align}
KKH^{-6}(pt,pt)=KSp(\IR^6)=KO(\IR^2)=\IZ_2,
\end{align}
while for $O1^+$ we have that
\begin{align}
KKH^{-2}(pt,pt)=KSp(\IR^2)=KO(\IR^6)=0.
\end{align}
This confirms that an orientifold classification in terms of branes rather than fields is easily gathered by KK-theory.

\section{D-branes in Orbifold singularities and KK-theory}

So far our main focus has been on the prescription of D-branes in
orientifolds. In this section we describe how to incorporate the
equivariant version $KK_{G}(X,Y)$ of the Kasparov KK-theory
bifunctor to the D$d$-brane classification scheme. The expected
group is $KK_{G}(X,Y)$ since the K-theory group classifying
D-branes in orbifold singularities is the equivariant group
$K_{G}(X)$ \cite{Witten:1998cd, GarciaCompean:1998rg}.

For simplicity we will concentrate in the case where the dimension
$q$ of the unstable D$q$-brane system is higher than the dimension
$p$ of the transverse space to the orbifold singularity. The
reader can extend the formulation for $q \leq p$ by following the
arguments in this section and section 3.3.

\subsection{Type IIB orbifolds and Equivariant $KK_{G}$-theory }

In order to describe orbifold singularities with equivariant
$KK_G$-theory we assume a group $G$ acting on the $(9-p)$
coordinates $(x^{p+1},...,x^{9})$ of spacetime in a Type IIB
string theory, i.e. the spacatime is

\begin{equation}
\qquad \qquad \qquad \qquad \qquad  \mathbb{R}^{p+1} \times
(\mathcal{M}^{9-p}/G). \label{orbi1}
\end{equation}
Let us concentrate on the case of flat spacetime by taking:
$\mathcal{M}^{9-p} = \mathbb{R}^{9-p}$. Then, the general form of
a subspace of spacetime is as follows:

\begin{equation}
\qquad \qquad \qquad \qquad \qquad \mathbb{R}^{\alpha,
\beta}=(\mathbb{R}^{\alpha}/G) \times \mathbb{R}^{\beta}.
\label{orbi2}
\end{equation}

In \cite{Gaberdiel:1999ch} it is shown that the K-theory group
classifying D-branes in orbifold singularities in Type IIB string
theory is the equivariant group $K_{G}(X),$ where $X$ is the
transverse space to the D$d$-brane with respect to the spacetime
(or an unstable system of $\hbox{D}9$-$\overline{\hbox{D}9}$
spacetime filling D-branes). The arguments explained in Sec. 2.1
for the IIB string theory can be applied here. Indeed, the
K$_{G}$-group classifying stable D-branes in the Type IIB orbifold
singularity with respect to an unstable D$q$-brane system is given
by K$_{G}^{q-1}(X),$ where $X$ is now the transverse space of the
stable D$d$-brane relative to the unstable D$q$-brane system, with
$p \leq q$.

Our goal is to classify all possible stable D$d$-branes in the
spacetime (\ref{orbi1}) by using equivariant KK$_{G}$-groups and
incorporating the unstable information of the D$q$-brane system
mentioned above.


Using the above remarks and appendix B.1, we claim that the group
classifying any D$d$-brane located in the orbifold singularity is
$KK_{G}^{q-1}(X,Y)$, where $Y$ is the portion of the spacetime
supporting the unstable D$q$-branes and $X$ is the transverse
space to $Y$ with coordinates $(x^{q+1}, \cdots ,x^{9})$ in the
whole spacetime. The present subsection will be devoted to prove
this claim.

It is worth to mention again the limiting cases. For D$d$-branes
extended totally outside the worldvolume of the unstable system,
we can take $Y$ as a point. Then $KK_{G}^{q-1}(X,pt)$ is the group
classifying D-branes extended along $X$. This is precisely the
K$_{G}$-homology of $X$ which classifies D-branes by their
worldvolume. Similarly, if the D-brane is extended completely
inside the unstable system, the group classifying stable D-branes
is the equivariant K-group $KK_{G}^{q-1}(pt,Y)=K_{G}^{q-1}(Y)$
classifying  in terms of the transverse space of the D-brane
relative to the unstable system.

Thus in the general case, given a D$d$-brane whose position lies
both inside and outside of the unstable D$q$-brane ambient, the
two entries of the KK$_{G}$-functor should be filled firstly by
the worldvolume $X$ of the portion of the D-brane outside the
unstable system. The second item $Y$ of codimension $m$
corresponds to the transverse space of the D$d$-brane in the
unstable D$q$-brane.

Then the spaces filling the KK$_{G}$-functor entries depend
strongly in the directions where the D$d$-brane is extended, but
not on its dimension $d$.


To be more specific, consider an unstable system of D$q$-branes
placed at the orbifold singularity and extended along
$(x^{0},...,x^{q})$ and place a D$d$-brane extended along
$(x^{0},...,x^{q-m},x^{q+1},...,x^{d+m})$ where the spacetime is
the orbifold defined above \footnote{If $d \geq q$ then $ m \leq
9-d$.}, with $q-m \leqq p \leqq q$ . Then the KK$_{G}$-theory
group classifying this system is

\begin{equation}
 \qquad \qquad  \qquad \qquad KK_{G}^{q-1}(\mathbb{R}^{d+m-q, 0},
  \mathbb{R}^{q-p,p-q+m}).
\label{pos2g}
\end{equation}
Using (\ref{kkbpg}) and assuming that $G$ acts by a spinor
representation we find that:

\begin{equation}
  \qquad KK_{G}^{q-1}(\mathbb{R}^{d+m-q,0},
  \mathbb{R}^{q-p,p-q+m})=
  K_{G}(\mathbb{R}^{9-p-(d+m-q), p-q+m}).
 \label{pos3g}
\end{equation}

At first sight the above equation depend on the dimension $q$ of
the unstable brane system, and this would rule out our proposal
because the D$q$-brane is an auxiliary device for the KK-theory
formalism, and the result should not depend on it. Then let us
argue that this result is indeed independent of $q$ and at the
same time we will see that our result is in full agreement with
\cite{Gaberdiel:1999ch}. Remember that from this reference for the
Type IIB orbifold with the group $\mathbb{Z}_{2}$ acting by
reflection on $n$ coordinates ($n=4$ mod $4$ in order to preserve
some supersymmetry), we say that a D$d$-brane is of type $(r,s)$,
where $d=r+s$, if it has $r+1$ Neumann directions with
$\mathbb{Z}_{2}$ acting trivially on them and $s$ Neumann
directions inverted by $\mathbb{Z}_{2}$. Then for a given
$(r+s)$-brane, the transverse space has dimension $9-(r+s)$; of
which $n-s$ directions are inverted under the action of
$\mathbb{Z}_{2}$. Then the K$_{G}$-theory group classifying
D$d$-branes in this orbifold is
\begin{equation}
\qquad \qquad \qquad \qquad \qquad
K_{\mathbb{Z}_{2}}(\mathbb{R}^{n-s,9-n-r}). \label{orbi4}
\end{equation}
This result is tested by computing these
K$_{\mathbb{Z}_{2}}$-groups and comparing the result with the
boundary state formalism, finding full agreement
\cite{Gaberdiel:1999ch}.

Though Eq. (\ref{orbi4}) is just for $\mathbb{Z}_{2}$, we will
prove that this result is valid for every group acting on the
spacetime by means of the spinor representation and then we will
argue that the action $\mathbb{Z}_{2}$ for which (\ref{orbi4}) is
valid  acts precisely in this way.  If we compare our original
system with the one described above, we find the following
correspondences:

\begin{equation}
\qquad \qquad \qquad n=9-p, \qquad d=r+s, \qquad s=d+m-q.
\label{orbi5}
\end{equation}
With these relations one can easily express (\ref{pos3g}) as

\begin{equation}
\qquad \qquad \qquad
 KK_{G}^{q-1}(\mathbb{R}^{s,0},\mathbb{R}^{q-p,p-q+m})=K_{G}(\mathbb{R}^{n-s,9-n-r}),
\label{orbi6}
\end{equation}
which is exactly (\ref{orbi4}) with a general group $G$ acting by
the spinor representation instead of $\mathbb{Z}_{2}$. In
(\ref{orbi4}) we only assume the existence of the
D$9$-$\overline{\hbox{D}9}$ unstable system and from (\ref{orbi6})
we see that for each D$q$-brane system our proposal is equivalent
to (\ref{orbi4}). Then we conclude that the KK$_{G}$-formalism is
independent of the D$q$-system.

Now we argue why the $\mathbb{Z}_{2}$-action assumed above is
spinor. In appendix D we mention that $G$ acts on $\mathbb{R}^{n}$
through the spinor representation if it acts by a group
homomorphism $G \mapsto Spin_{n}$. By this we mean a homomorphism
$\alpha :G \mapsto Spin_{n}$ such that, when composed with the
natural action of $Spin_{n}$ on $\mathbb{R}^{n}$ $(x \mapsto
\gamma x \gamma ^{-1}, \quad x \in \mathbb{R}^{n}, \quad \gamma
\in Spin_{n})$ we get a representation (which induces an action)
of $G$ on $\mathbb{R}^{n}$.

Consider a D$4$-brane ($d=4$) such that $r=1$ and $s=3$. Then we
can think of the orbifold $\mathbb{Z}_{2}$-action $x \sim -x$,
with $x\in\mathbb{R}^{3}$ as a $\pi$-rotation around some rotation
axis in $\mathbb{R}^{3}$; but we know that each rotation in
$\mathbb{R}^{3}$ can be generated by $SU(2) \simeq Spin_{3}$
acting on $\mathbb{R}^{3}$ through Pauli matrices. In our
particular situation, the homomorphism assigning to $-1 \in
\mathbb{Z}_{2}$ the $\pi$-rotation $U$, such that
$$
x \mapsto UxU^{-1}=-x, \qquad x \in \mathbb{R}^{3}
$$ does the
job. Therefore we can see in this particular example how the
$\mathbb{Z}_{2}$ orbifold action considered in
\cite{Gaberdiel:1999ch} fits in our formalism.

Though it is not easy to find the homomorphism $G \mapsto
Spin_{n}$ for higher values of $n$, the arguments given above
generalize to any $n$ because $Spin_{n}$ is the double covering of
$SO(n)$, and consequently, to each $SO(n)$ rotation always
correspond at least one element in $Spin_{n}$.

We have generalized \cite{Gaberdiel:1999ch} (at least for the case
of flat noncompact orbifolds) because our formalism applies to any
group action on the spacetime which can be described as a rotation
around some axis. Our result also includes some of the examples
studied recently in \cite{Kriz:2007ik}, where the orbifold actions
are rotations around some axis of the spacetime. In particular, we
generalized the flat orbifolds in \cite{Kriz:2007ik} of the form
$\mathbb{Z}_{k}$ for any $k \in \mathbb{N}$ and $\mathbb{Z}_{k}
\times ... \times \mathbb{Z}_{k}$ (without discrete torsion).

Now we consider an example discussed in \cite{Kriz:2007ik}. This
is the orbifold $\mathbb{C}^{3} / \mathbb{Z}_{3}$, with spacetime
of the form $ \mathbb{R}^{4}
 \times \mathbb{C}^{3} / \mathbb{Z}_{3}$, where $(x^{0},
 x^{1},x^{2},x^{9})$ are the coordinates in which $
 \mathbb{Z}_{3}$ acts trivially and $ z^{i}={2^{- \frac{1}{2}}}(x^{2i+1} +
 x^{2i+2})$ $i=1,2,3$ are the coordinates where the generator $g$
 of $G$ acts in the form
\begin{equation}
\qquad g(z^{1},z^{2},z^{3}) \rightarrow (\exp(2 \pi i
\upsilon_{1}) z^{1}, \exp(2 \pi i \upsilon_{2}) z^{2}, \exp(2 \pi
i \upsilon_{3}) z^{3}),
 \label{orbi8}
\end{equation}
where $(\upsilon_{1},\upsilon_{2},\upsilon_{3})=(\frac{1}{3},
\frac{1}{3}, - \frac{2}{3})$. The action \eqref{orbi8} is clearly
a rotation in $\mathbb{C}^{3}$. Therefore the action of
$\mathbb{Z}_{3}$ on any $D$-brane with $s$ Neumann coordinates in
$\mathbb{C}^{3} / \mathbb{Z}_{3}$ can be seen as a rotation of the
$s$-coordinates and hence as a spin representation of
$\mathbb{Z}_{3}$ in $Spin_{s}$ and hence this example is also
included in our formalism.

Now we describe the gauge theory living in the unstable $Dq$-brane
system \cite{Douglas:1996sw}. Following \cite{Asakawa:2002nv}, we
focus on the stable D-branes which are outside the worldvolume $Y$
of the unstable D$q$-brane system and consequently $Y$ can be set
to be a point. In our case we need to take into account the images
of this point under the action of the group; thus we set $Y$ to be
the space of $t$-points, where $t$ is the cardinality of $G$. The
$KK_{G}$-theory group classifying the above D-branes with respect
the D$q$-unstable system is given by
$$
KK_{G}^{q-1}(X, \{t \quad {\rm points}\})=KK_{G}(C_{0}(X), C(\{t
\quad {\rm points}\}) \otimes \mathbb{C}l^{9-q}
$$
 $$
 =KK_{G}(C_{0}(X),(\oplus_{i=1}^{t} \mathbb{C}) \otimes
 \mathbb{C}l^{9-q})= KK_{G}(C_{0}(X),\oplus_{i=1}^{t} (\mathbb{C}
\otimes \mathbb{C}l^{9-q})) $$
\begin{equation}
\qquad \qquad = \bigoplus_{i=1}^{t}KK_{G}(C_{0}(X), \mathbb{C}
\otimes \mathbb{C}l^{9-q}). \label{orbi7}
\end{equation}
Thus we can associate a ``gauge group'' for each direct summand in
\eqref{orbi7}. If we assemble these gauge groups in a block
diagonal matrix, we get a matrix $M$ with $t$ blocks  and such a
matrix belongs to the algebra $\textbf{B}(C_{0}(t \quad {\rm
points})^{\infty}) \otimes \mathbb{C}l^{9-q}$ of adjointable
operators on $ C_{0}(t \quad {\rm points})^{\infty} \otimes
\mathbb{C}l^{9-q}$, which is the appropriate Hilbert module for
describing $KK_{G}^{q-1}(X, \{t \quad {\rm points} \}).$ Moreover,
$M$ represents the gauge group  of the low energy effective field
theory on the D$q$-brane worldvolume, which is of the form (for a
finite number of branes) $\prod_{i=1}^{t}U(N_{i})$ or
$\prod_{i=1}^{t}(U(N_{i}) \times U(N_{i}))$, where $R=
\oplus_{i=1}^{t}N_{i}r_{i}$ is the representation of $G$ on the
Chan-Paton factors and $r_{i}$ are the irreps of $G$. Of course,
each block in $M$ is infinite because in order the KK-theory make
sense we must assume the presence of an infinite number of
D$q$-branes \cite{Asakawa:2001vm}.

If we take for instance, $q=7$ then  following Ref.
\cite{Asakawa:2002nv} and the preceding section, the gauge group
associated to each of the factors in
$\bigoplus_{i=1}^{t}KK_{G}(C_{0}(X), \mathbb{C} \otimes
\mathbb{C}l^{9-q}$ is determined by $[\mathbb{C}l^{2}_{even}]=
\mathbb{C} \oplus \mathbb{C}$ which corresponds to $U(\infty)
\otimes U(\infty)$. Thus the gauge group of the unstable
D$7$-brane system in the orbifold singularity is (as expected)
given by $\prod_{i=1}^{t}(U(\infty) \otimes U(\infty))$.

\section{Final Remarks}

In this paper, we have extended to Type IIB orbifold and
O$p^{-}$-orientifold backgrounds the KK-theory formalism proposed
in \cite{Asakawa:2001vm, Asakawa:2002nv} for Type IIB and Type I
string theory respectively.

In particular, for the orientifold case, we considered $Op^-$-planes with $p=1,5,9$, for which
the presented formalism naturally incorporates stable D-branes
intersecting the orientifold planes, generalizing in this sense the proposal in \cite{Gukov:1999yn} for the mentioned cases. This is achieved by constructing D-branes from unstable D$q$-branes in which the final D-brane has internal and external coordinates with respect to the D$q$-brane. In this sense, the internal coordinates are identified with the space $Y$ and the external ones with the space $X$, where $X$ and $Y$ are the entrances in the KKR-theory bifunctor $KKR(X,Y)$.

Specifically we propose that D$d$-branes intersecting
$Op^-$-planes are given by the groups in Eqs.(\ref{Eq:KKR1}) and
(\ref{Eq:KKR2})
\begin{eqnarray}
KKR(\IR^{d-s-r,0}, \IR^{(9-q)+(q-p-r), p-s})&=&KKR^{1-q}(\IR^{d-s-r,0},\IR^{q-p-r,p-s})\quad \hbox{for $p<q$},\nonumber\\
KKR(\IR^{d-s, r}, \IR^{9-p, p-q+(q+r-s)})&=&KKR^{9-2p+q}(\IR^{d-s,r},\IR^{0,q+r-s})\quad \hbox{for $p>q$}.\nonumber
\end{eqnarray}
In order to show that these groups correctly classify the
corresponding configurations of D-branes and orientifolds, we also
compute, by extensive use of the Clifford algebras and the
structures defined on them, the gauge group and transformation
properties of the effective fields living in the worldvolume of
the unstable D$q$-branes. The transformation properties of the
tachyon and scalar fields of this unstable systems are read from
the {\it fixed point Clifford algebra}. This algebra consists of
Clifford generators invariant under the involution, determined in
turn by the action of the corresponding orientifold plane. The set
of algebras related to different configurations is listed in Table
1 in text. In all cases we found perfect agreement with Type I
T-dual versions, as reported in \cite{Asakawa:2002nv}. This shows
that Clifford algebras contain relevant information about
stability, RR charge and construction of D-branes in general
backgrounds.

However, although this formalism seems powerful enough, the
mathematical information in literature concerning other physical
relevant cases, as positive RR charged orientifolds, is limited.
Working out with expected physical properties instead, we have
proposed some KK-theory groups related to the mentioned cases. In
particular we have proposed some versions of KK-theory (KKH and
KKSp) based on the existence of consistent string theories with
D-branes carrying quaternionic and symplectic Chan-Paton bundles.
Moreover, we propose, based on their respective K-theories some
simple properties of these bifunctors. We also give some clues on
how the appropiate structures should be incorporated on the
Kasparov modules entering the KKH-theory definition. Similar
arguments should apply to KKSp-theory.

As a matter of probe, we have applied this formalism, including
the proposal on positive RR charged orientifolds, to elucidate the
existence of the so called {\it exotic} orientifold planes. These
planes have been classified by K-theory, but in terms of RR
fields. A brane realization of exotic planes reveals a
configuration of brane and orientifold planes, for which it is
possible to apply the present formalism. For the considered cases,
we have found that a brane classification of this planes is
possible by means of KK-theory. The results are also in agreement
with the RR field classification.

In the orbifold case we reproduce the proposal in
\cite{Gaberdiel:1999ch} in terms of equivariant
$\hbox{K}_{G}$-theory for a $\mathbb{Z}_{2}$-orbifold. Moreover,
we argued that this prescription is valid for any $G$-action by
means of the spinor representation. In this way our formalism
includes every $G$-action that can be seen as a space-time
rotation; including in particular the Type IIB examples considered
in \cite{Kriz:2007ik} of flat orbifolds without discrete torsion
(we include an explicit example of the orbifold $\mathbb{C}^{3} /
\mathbb{Z}_{3}$ discussed in this reference and show that
$\mathbb{Z}_{3}$ acts on $\mathbb{C}^{3}$ by the spinor
representation), where the explicit $\hbox{K}_{G}$-groups (and
hence $\hbox{KK}_{G}$-groups) are calculated. We also recover the
gauge theory in the unstable D$q$-brane systems of Type IIB string
theory orbifold spacetimes.


In \cite{Moore:1999gb} the K-theory formalism is incorporated to
the classification of fluxes in Type IIB string theory which are
not sourced by D-branes. It is then natural to incorporate the
KK-theory formalism for the classification of these fluxes. Some
research in this direction was addressed in \cite{Reis:2006th,
Brodzki:2006fi}

 In \cite{Bergman:1999ta} T-duality is explained in terms of certain
isomorphisms of relative K-theory for spacetime compactifications
in $\bf{T}^n$. So, compactifying the spacetime, amounts to define
``relative KK-theory'' \cite{higson} and the incorporation of
T-duality would imply some isomorphisms in the corresponding
``relative KK-groups''. Some considerations about T-duality and
KK-theory has been discussed in another context in
\cite{Brodzki:2006fi, Brodzki:2007hg} (for a recent review see
\cite{Szabo:2008hx}). Finally, in analogy with \cite{Kapustin:1999di} it
should be interesting to incorporate a topologically non-trivial
B-field background to the KK-theory classification of D-branes,
leading to a twisted KK-theory.

\begin{center}
{\bf Acknowledgments}
\end{center}

It is a pleasure to thank A. Hanany and S. Sugimoto for useful
explanations. W.H.S wish to thank the Centro de Investigaci\'on y
de Estudios Avanzados, Unidad Monterrey for its hospitality during
the realization of this work. The work of W.H.S. was supported by
a CONACYT grant 171089. This work was partially supported by
CONACyT's grants 45713-F and 60209.

\appendix

\section{Complex KK-theory}

We start by defining a Kasparov module\footnote{Though there are
several approaches to Kasparov modules \cite{kaspa:1981wl,
Blackadar:1998oa, Schroder:2001ms}, we will use the Fredholm
picture which fills out our requirements for physical
interpretations.}. Let $(A,B)$ be a pair of trivially graded,
separable, unital and complex $C^{*}$-algebras. An \textit{odd
Kasparov $A$-$B$ module} is a triple $({\cal H}_{B},\phi,T)$,
where

\begin{itemize}

    \item ${\cal H}_{B}=B^{\infty}$ is the Hilbert $B$-module defined
    as follows:

    $$B^{\infty}=\{(x_{k})\in\prod_{n=1}^{\infty} B \mid \sum_{k} x_{k}^{*}x_{k}\
    converges\ in\ B\}.$$

    \item $\phi:A \rightarrow \textbf{B}({\cal H}_{B})$ \footnote{If $E$ is any Hilbert $B$-module for
    a $C^{*}$-algebra $B$, we will denote as \textbf{B}($E$) the set of adjointable operators i.e. the operators
    $T:E \rightarrow E$ such that there exist an operator $T^\dagger :E \rightarrow E$ with $(Ta,b)=(a,T^\dagger b)$ for all $a,b \in$ $E$, and $(a,b)$ is the $B$-valued inner product of $
\mathcal{H}_{B}$ as a Hilbert $B$-module.} is a unital
    $*$-homomorphism.

    \item $T \in \textbf{B}({\cal H}_{B})$ is a self-adjoint operator
    such that

\begin{equation}
             T^{2}-1, \quad [T,\phi (a)] \in {\cal K}({\cal
             H}_{B})=B \otimes {\cal K} \quad {\rm{for \; all}} \ a  \in A,
             \label{tachyon}
 \end{equation}

 \noindent where ${\cal K}({\cal H}_{B})$ is defined such that any pair of elements, $x, y \in {\cal H}_{B}$, gives rise to a
 map $\Theta_{x,y}:{\cal H}_{B} \rightarrow {\cal H}_{B}$ given by
 $\Theta_{x,y}(z)=x(y,z),$ for all $z \in {\cal H}_{B}.$ Then ${\cal K}({\cal
 H}_{B})$ is the closed linear span of $\{ \Theta_{x,y} : x,y \in
 {\cal H}_{B} \}$ and it is a closed two sided ideal in $\textbf{B}({\cal
 H}_{B})$. Note that when $B$ is the field of complex numbers, then ${\cal K}({\cal
 H}_{B})$ is identified with the space of compact operators on ${\cal H}_{B}$
 (denoted ${\cal K}$), and ${\cal H}_{B}$ is identified with the space of square summable sequences
 in the complex numbers.

\end{itemize}

An \textit{odd Kasparov $A$-$B$ module} is called
\textit{degenerate} if

\begin{equation}
      \qquad \qquad \qquad           T^{2}-1=[T,\phi (a)]=0 \quad {\rm{for \; all}} \ a \in A.
           \label{degenerate}
\end{equation}

Now, we define some relations on the set of \textit{odd Kasparov
$A$-$B$ modules}:

\begin{itemize}

    \item Two triples $({\cal H}_{B}, \phi_{0}, T_{0})$ and $({\cal
     H}_{B}, \phi_{1}, T_{1})$ are called \textit{unitarily equivalent}
     if there exists an unitary operator  $U \in \textbf{B} ({\cal H}_B)$
     with $T_{0} = U^{*} T_{1} U$ and $\phi_{0}(a)=U^{*} \phi_{1}(a)U$
     for all $a \in A.$

    \item Let $({\cal H}_{B}, \phi_{i}, T_{i})$ be \textit{odd Kasparov $A$-$B$ modules}
    for $i=0,1$; let $(E, \phi, T)$ be an \textit{odd Kasparov $A$-$B \otimes C[0,1]$
    module} and let $\textit{f}_{t}:B \otimes C[0,1] \rightarrow B$
    denote the evaluation map $\textit{f}_{t}(g)=g(t)$. Then
    $({\cal H}_{B}, \phi_{0}, T_{0})$ and $({\cal H}_{B}, \phi_{1},
    T_{1})$ are called \textit{homotopic} and $(E, \phi, T)$ is
    called a \textit{homotopy} if $(E \otimes_{\textit{f}_{i}} B, \textit{f}_{i} \circ \phi,
    \textit{f}_{i_{*}}(T))$ is unitarily equivalent to $({\cal
    H}_{B}, \phi_{i}, T_{i}), \ i=0,1$, where
    $\textit{f}_{i_{*}}(T)(a):=\textit{f}_{i}(T(a)).$

    \item If $E=C([0,1], {\cal H}_{B})$ and for all $a \in A$ the induced maps $t
    \rightarrow T_{t}, t \rightarrow \phi_{t}(a)$
    are strongly $*$-continuous, then $(E, \phi, T)$ is called a
    \textit{standard homotopy}. When in
    addition $\phi_{t}$ is constant and $T_{t}$ is norm continuous
    then we say that $(E, \phi, T)$ is an \textit{operator homotopy}.

    \item In the definition of the group $KK^{1}(A,B)$, two \textit{odd Kasparov
    modules} $({\cal H}, \phi_{i}, T_{i})$, $\quad i=0,1$ are defined to
    be equivalent (and denoted $\sim_{oh}$) if there are degenerate
    Kasparov modules $({\cal H}_{B}, \phi_{i}',T_{i}')$, $\quad
    i=0,1$ such that $({\cal H}_{B} \oplus {\cal H}_{B}, \phi_{i}
    \oplus \phi_{i}', T_{i} \oplus T_{i}')$, $\quad i=0,1$ are
    operator homotopic up to unitary equivalence. Then
    $KK^{1}(A,B)$ is the set of equivalence classes of odd
    Kasparov modules under $\sim_{oh}$.

    \end{itemize}

    The other $KK$-group $KK(A,B) \equiv KK^{0}(A,B)$ is the set of
    equivalence classes of $\mathbb{Z}_{2}$-graded triples $(\widehat{{\cal
    H}}, \widehat{\phi}, F)$, called \textit{even Kasparov $A$-$B$ modules},
    with

\begin{equation}
    \qquad \qquad \widehat{{\cal H}}={\cal H}_{0} \oplus {\cal H}_{1}, \qquad
    \widehat{\phi}=diag(\phi_{0}, \phi_{1}), \qquad \mathit{F}=
    \left(\begin{array}{cc} 0 & T^\dagger \\  T & 0 \end{array}
    \right),
\label{grading}
\end{equation}

    \noindent where ${\cal H}_{i} \ (i=0,1)$ are Hilbert $B$-modules,
    $\phi_{i}:A \rightarrow B({\cal H}_{i})$ is a unital
    $*$-homomorphism for $i=0,1$ and $T \in \textbf{B}({\cal
    H}_{0}, {\cal H}_{1})$ is an adjointable operator such that

\begin{equation}
   \qquad  T^{\dagger}T -1, \qquad TT^\dagger-1, \quad T \phi_{0}(a)- \phi_{1}(a)T \in B
    \otimes {\cal K} \quad {\rm{for \; all}} \ a  \in A.
\label{gradedtachyon}
\end{equation}

    The grading is induced by the standard even grading operator
    $diag(1,-1)$ where we identify $\textbf{B}(\hat{{\cal
    H}}_{B})=M_{2}(M(B \otimes {\cal K}))=M(B \otimes {\cal K})$,
    where $M(B \otimes {\cal K})$ is the multiplier
    algebra of $B \otimes {\cal K}$ and $M_{2}(A)$ is the
    $C^{*}$-algebra of $2 \times 2$ matrices with entries in the
    $C^{*}$-algebra $A$.

    The group $KK^{0}(A,B)$ is defined as the set of equivalence
    classes of even Kasparov modules with the equivalence relation
    $\sim_{oh}$ defined above.

    It can be proved that $KK^{1}(A,B)=KK^{0}(A,B \otimes
    \mathbb{C}l^{1})$ with $\mathbb{C}l^{1}$ being the complex Clifford
    algebra generated by $\{1, e_{1}\}$, where $e_{1}^{2}=1$. This is
    the approach we will take for the definitions of the real,
    Real and equivariant $KK$-groups for real, Real $C^{*}$-algebras and $G$-algebras. In general,
    one can define higher $KK$-groups as $KK^{n}(A,B)=KK(A,B \otimes
    \mathbb{C}l^{n})$, but periodicity mod 2 tells us that the only
    $KK$-theory groups are $KK^{0}$ and $KK^{1}$.

\section{Orthogonal (real) KK-theory}

The real $KKO$-group is defined similarly as above, but
substituting complex objects by real ones (real $C^{*}$-algebras,
real Hilbert $B$-modules, etc.). In other words, the structures
are defined over the field of the real numbers instead of the
complex numbers field.

To define the higher real $KKO$-groups we need to define a real
$C^{*}$-algebra structure in the real Clifford algebra
$\textsl{Cl}^{n,m}$, where $\textsl{Cl}^{n,m}$ is generated (as an
algebra over $\mathbb{R}$) by $\{e_{i} \in \mathbb{R}^{n+m}
\}_{i=1,...,n+m}$ with the relations

\begin{eqnarray}
 \qquad \qquad \qquad \qquad \qquad e_{i}e_{j}+ e_{j}e_{i} = 0 \quad (i \neq j),
 \nonumber
 \\e_{i}^{2} = -1\quad(i=1,...,n),
 \\e_{i}^{2}  =  1 \quad (i=n+1,...,n+m).  \nonumber
 \label{clifford1}
 \end{eqnarray}


The $C^{*}$-algebra involution is defined on the generators
$\{e_{i} \in \mathbb{R}^{n+m} \}$ as follows:

\begin{eqnarray}
\qquad \qquad \qquad \qquad \qquad e_{i}^{*}=-e_{i} \quad
(i=1,...n),
\nonumber  
\\e_{i}^{*}=e_{i} \quad (i=n+1,...,n+m), \label{ccliff}
\\(e_{1}\cdots e_{l})^{*}=e_{l}^{*} \cdots e_{1}^{*}, \nonumber
\end{eqnarray}



\noindent and extending it by linearity.

It can be shown \cite{kaspa:1981wl, Schroder:2001ms} that $KKO(A
\otimes \textsl{Cl}^{n,m}, B \otimes \textsl{Cl}^{r,s})$ depends
only on $(m-n)-(s-r)$, so we can define with no ambiguity:

\begin{equation}
\qquad \qquad KKO_{m-n+r-s}(A,B)=KKO(A \otimes \textsl{Cl}^{n,m},
B \otimes \textsl{Cl}^{r,s}). \label{kko}
\end{equation}

 Thus, for $n \in \mathbb{Z}$ we define

 \begin{equation}
\qquad $$KKO^{-n}(A,B)=KKO_{n}(A,B)= \left\{
\begin{array}{cc} KKO(A,B \otimes \textsl{Cl}^{n,0}) & n > 0 \\
KKO(A,B \otimes \textsl{Cl}^{0,-n}) & n < 0  \end{array}
\right\}.$$ \label{kko2}
\end{equation}

\noindent The $KKO_{n}$-groups are periodic mod 8.

Both, complex and real $KK$-theories share the following crucial
property called Bott Periodicity:

\begin{equation}
\qquad $$KKO^{k}(X,Y)$$=$$KKO^{k-n}(X\times
\mathbb{R}^{n},Y)$$=$$KKO^{k+m}(X,Y \times \mathbb{R}^{m}),$$
\label{kkbp}
\end{equation}

\noindent where $\mathbb{R}^{n}$ stands for
$C_{0}(\mathbb{R}^{n})$. In general, for any locally compact
topological spaces $X$ and $Y$, we denote $KKO^{n}(X,Y) \equiv
KKO^{n}(C_{0}(X),C_{0}(Y))$, where $C_{0}(X)$ $(C_{0}(Y))$ is the
$C^{*}$-algebra of continuous real (or complex when we are dealing
with KK-groups) valued functions on $X$ $(Y)$, vanishing at
infinity.

\subsection{$KK$-Theory applied to D-Branes}

Suppose we have a spacetime of the form $X \times Y$ (with
$dimY=q+1$) and an unstable system of an infinite number of
D$q$-branes extended on $Y$, in Type II string theory. It was
proposed in \cite{Asakawa:2001vm} that the solitonic
configurations (which turn out to be D-branes) of this system are
classified by the complex KK-groups: The $KK^{1}(X,Y)$-group for
non-BPS D$q$-branes and $KK^{0}(X,Y)$-group for a D$q$-
$\overline{\hbox{D}q}$ system ($\overline{\hbox{D}q}$ denotes an
anti D$q$-brane) of stable D$q$ branes. The grading for the even
Kasparov modules described in the definition of the KK$^{0}$-group
is associated with the D$q$- $\overline{\hbox{D}q}$-branes.


 Now, we will review the way KKO-groups are applied to classify D-branes in Type I
string theory.

In \cite{Bergman:2000tm} it is argued that the K-theory group
classifying D$d$-brane charges inside the worldvolume of an
unstable D$q$-brane system in Type I string theory is the real
K-theory group $KO^{q-1}(Y)$, where $Y$ is the worldvolume
manifold of the unstable system. The proposal in
\cite{Asakawa:2002nv} is that the group that correctly classifies
D-branes stretched along both longitudinal and transverse
directions to the unstable D$q$-brane system is $KKO^{q-1}(X,Y)$
where $X$ and $Y$ have the same meaning that in the complex case.

The elements of $KKO^{0}(X,Y)$ can be interpreted in the same way
as for Type II string theories in terms of D$q$-
$\overline{\hbox{D}q}$-brane system (for $q=0,9$) wrapped in $Y$.

In general $KKO(C_{0}(X), C_{0}(Y) \otimes
\textsl{Cl}^{n,m})$ consists of equivalence classes of triples 
$(\widehat{{\cal H}}, \widehat{\phi}, F)$ where ${\cal
\widehat{H}}=C_{0}(Y)^{\infty} \otimes \textsl{Cl}^{n,m}, \
\widehat{\phi}:C_{0}(X) \rightarrow \textbf{B}({\cal
\widehat{H}})$ is a $*$-homomorphism, and $F$ is a self-adjoint
operator in $\textbf{B}({\cal
\widehat{H}})=\textbf{B}(C_{0}(Y)^{\infty} \otimes
\textsl{Cl}^{n,m})$. We have the additional requirement that
$\widehat{\phi}(a) (a \in C_{0}(X))$ be even and $F$ odd with
respect to the $\mathbb{Z}_{2}$-grading. In this way, we can
write:

\begin{equation}
\qquad \qquad  F= \sum_{\textit{v}_{r} \in
\textsl{Cl}^{n,m}_{odd}} T_{r} \otimes \textit{v}_{r}, \qquad
\widehat{\phi}(a)= \sum_{\textit{w}_{r} \in
\textsl{Cl}^{n,m}_{even}} \Phi_{r} \otimes \textit{w}_{r},
\label{tachyon2}
\end{equation}

 \noindent where $T_{r}, \Phi_{r} \in \textbf{B}(C_{0}(Y))$, $\textit{w}_{r}$ and $\textit{v}_{r}$ span the sets of even
an odd elements in $\textit{Cl}^{n,m}$ denoted as
$\textsl{Cl}^{n,m}_{even}$ and $\textsl{Cl}^{n,m}_{odd}.$

\subsection{Field content in unstable Type I non-BPS branes}
As it was studied in \cite{Asakawa:2002nv}, the field content
representation of an unstable non-BPS D$q$-brane in Type I theory
can be elucidated from the real Clifford algebra since in this
case, as can be seen from B.1, the tachyon and scalar fields
satisfy some requirements. The tachyon field is an odd
self-adjoint operator, while $\phi$ is an even self-adjoint map.
This fixes their representations under the gauge transformation,
which is an even unitary transformation on the Hilbert space
${\cal H}$. For completness we reproduce the results obtained in
\cite{Asakawa:2002nv} in Table \ref{Tabla:japon}.

\begin{table}
\begin{center}
\begin{tabular}{|c||c|c|c|c|}
\hline
D$q$&$Cl^{n,m}$&$\phi$&$T$&Gauge group\\
\hline\hline
D(-1)&$Cl^{2,0}$&$adj.$&${\Yasymm}$&$U(\infty)$\\
D7&&&&\\
\hline
D0&$Cl^{1,0}$&${\Ysymm}$&${\Yasymm}$&$O(\infty)$\\
D8&&&&\\
\hline
D1&$Cl^{1,1}$&(1,$\Ysymm$),($\Ysymm$,1)&($\fund$, $\antifund$)&$O(\infty)\times O(\infty)$\\
D9&&&&\\
\hline
D2&$Cl^{0,1}$&${\Ysymm}$&${\Ysymm}$&$O(\infty)$\\
\hline
D3&$Cl^{0,2}$&$adj.$&${\Ysymm}$&$U(\infty)$\\
\hline
D4&$Cl^{0,3}$&$\Yasymm$&${\Ysymm}$&$Sp(\infty)$\\
\hline
D5&$Cl^{4,0}$&(1,$\Yasymm$),($\Yasymm$,1)&($\fund$, $\antifund$)&$Sp(\infty)\times Sp(\infty)$\\
\hline
D6&$Cl^{3,0}$&$\Yasymm$&$\Yasymm$&$Sp(\infty)$\\
\hline
\end{tabular}
\caption{Field content of unstable non-BPS Dq-branes in type I
theory and the relevant real Clifford algebra, as obtained in
Reference \cite{Asakawa:2002nv}.} \label{Tabla:japon}
\end{center}
\end{table}

For D$q$-branes in orientifold backgrounds like those studied in
this paper ($O1^-$ and $O5^-$) it suffices to compute the fixed
point algebra $(^p\IC l^{n,0})_\text{fix}$ or $(^p\IC
l^{0,n})_\text{fix}$. These fix point algebras are in general of
the form $Cl^{n,0}$ or $Cl^{0,n}$ (with
the only exception of algebras related to non-BPS D1 and D9 branes). This algebra fixes
in turn the field content and gauge group for each case. This was
explicity done in \cite{Asakawa:2002nv} which results are
summarized in Table \ref{Tabla:japon}.

\section{Clifford Algebras and Real KKR-Theory }

In this appendix we will describe the additional structure in the
Clifford algebras which is necessary for the definition of the KKR
bifunctor.

\subsection{Even and odd parts of the Real Clifford algebra}
In the Clifford algebra $Cl^{n,m}$ there is a natural grading
induced by the map $\alpha : Cl^{n,m} \mapsto Cl^{n,m}$ acting on
the generators like $\alpha (e_{i})=-e_{i}$ for all $i=1,...,n+m.$
Then, $\alpha$ is extended to the whole Clifford algebra by
linearity. In this way, the real Clifford algebra splits  in even
and odd parts, defined as the eigenspaces with eigenvalues 1 and
-1 respectively, i.e.

\begin{align}
Cl^{n,m}=(Cl^{n,m})_\text{even} \oplus (Cl^{n,m})_\text{odd},
\end{align}

 \noindent where $a \in (Cl^{n,m})_\text{even}$ if $\alpha(a)=a$ and
$a \in(Cl^{n,m})_\text{odd}$ if $\alpha(a)=-a.$

A general element of $Cl^{n,m}$ can be written as

\begin{align}
\alpha = a^{i_1}e_{i_1} + a^{i_1 i_2}e_{i_1i_2}+\cdots +a^{1\cdots
(n+m)}e_{1\cdots (n+m)}=
\sum_{k=1}^{n+m}\quad\!\!\!\!\!\sum_{i_1<i_2<\cdots <i_k}
a^{i_1\cdots i_k}e_{i_1\cdots i_k},
\end{align}

\noindent where $e_{i_1\dots i_k}\equiv e_{i_1}\cdot e_{i_2}\cdot
{\dots} \cdot e_{i_k}.$

It is easy to show that the even and odd parts can be expressed in
the following way:

\begin{eqnarray}
(Cl^{n,m})_\text{even}&=& \left\{\alpha \in Cl^{n,0}|\alpha=a^0 +a^{i_1i_2}e_{i_1i_2}+a^{i_1i_2i_3i_4}e_{i_1i_2i_3i_4}+\dots \right\},\nonumber\\
(Cl^{n,m})_\text{odd}&=& \left\{\alpha \in
Cl^{n,0}|\alpha=a^{i_1}e_{i_1}+a^{i_1i_2i_3}e_{i_1i_2i_3}+\dots
\right\}.
\end{eqnarray}
From now, we will concentrate on the algebras $Cl^{n,0}$
\footnote{Notice that for the complex Clifford algebra $\IC
l^{n}$, we have $\IC l^n=Cl^{n,0}\otimes \IC \cong Cl^{0,n}\otimes
\IC$. Then all expressions and facts in this appendix are valid
for the analog ones for $Cl^{0,n}$.}.

The grading in the Clifford algebra $Cl^{n,0}$ induce a grading in
the complex Clifford algebra $\IC l^n$ as follows:

\begin{eqnarray}
\qquad \qquad \qquad \qquad \IC l^n&=&Cl^{n,0}\otimes \IC \nonumber\\
&=&\left((Cl^{n,0})_\text{even}\oplus(Cl^{n,0})_\text{odd}\right)\otimes\IC \nonumber\\
&=&((Cl^{n,0})_\text{even} \otimes \IC) \oplus
((Cl^{n,0})_\text{odd}
\otimes \IC) \nonumber\\
 &\equiv&\IC l^{n,0}_\text{even}\oplus \IC l^{n,0}_\text{odd}.
\end{eqnarray}
Hence, an element $a$ in the even part of $\IC l^n\equiv\IC
l^{n,0}$ is just written as $\alpha\oplus(\beta\otimes i)$ with
$\alpha, \beta \in Cl^{n,0}_\text{even}$, and a similar expression
for the odd part.

\subsection{Fixed point algebra}
An element $a$ in the complexified Clifford algebra $\IC
l^{n}=Cl^{n,0}\otimes \IC$ is given by
\begin{align}
a= \alpha\oplus (\beta \otimes i),
\end{align}
where $\alpha$ and $\beta$ are elements of the real Clifford
algebra $Cl^{n,0}.$

Suppose that there is a Real involution defined on $\IC l^{n}$,
such that it is now a Real algebra denoted $~^p \IC l^{n,0}$ (as
explained in Sec.~3.2, the involution is given by the action of
the orientifold plane on the generators of the algebra and
extended by linearity).

On the other hand we have

\begin{align}
^p\IC l^{n,0}=~^p \left(Cl^{n,0}\otimes \IC\right)=~^pCl^{n,0}
\otimes \overline{\IC},
\end{align}

\noindent where $\overline{\IC}$ denotes the field of complex
numbers with Real involution defined by usual complex conjugation
and $~^pCl^{n,0}$ denotes the Clifford algebra $Cl^{n,0}$ with
some Real involution (again, this involution is determined by the
orientifold plane on the generators of the algebra and extended by
linearity). Then it is enough to select the proper involution on
the generators of $Cl^{n,0}$ to know the involution on $^p\IC
l^{n,0}$.

 The fixed point algebra of $^p\IC l^{n,0}$ is defined as
the set of elements in $^p\IC l^{n,0}$ which are invariant under
the involution\footnote{This definition applies to any algebra
with some Real involution.}, i.e.
\begin{align}
\left(~^p \IC l^{n,0}\right)_\text{fix}=\left\{a\in ~^p\IC l^{n,0}
|~a=\overline{a} \equiv {\cal I}_{9-p}(a)\right\},
\end{align}
where ${\cal I}_{9-p}(i)=-i$. Hence an element of the fixed point
algebra must satisfy the following constraint
\begin{align}
\overline\alpha \oplus (\overline\beta\otimes
\overline{i})=\overline\alpha \oplus (\overline\beta\otimes
(-i))=\overline\alpha \oplus ((-\overline\beta)\otimes i)=
\alpha\oplus (\beta\otimes i). \label{fixelement}
\end{align}

As a trivial example, consider $p=9$. Then for any unstable
D$q$-brane system\footnote{For simplicity, suppose $q>2$.}, we
have $q \leq p$. Following the criteria explained in Sec.~4.1, we
define the involution to be the trivial one in each generator of
the relevant Clifford algebra $Cl^{0,q-1}$. Then, by
(\ref{fixelement}) we identify

\begin{align}
\left(^9\IC l^{0,q-1}\right)_\text{fix}=Cl^{0,q-1}.
\end{align}


This is expected since in an $O9^-$-plane background, i.e. in Type
I theory, the whole nine-dimensional space is a fixed point under
the orientifold involution and hence, D-branes are characterized
by orthogonal Clifford algebra, as shown in \cite{Asakawa:2002nv}.


As explained in Sec.~4, the tachyon, the scalar fields and the
gauge transformation on the unstable D$q$-brane system all belong
to the fixed point algebra and to some of the even or odd parts of
$^9\IC l^{n,0}$ for some $n$. It turns out that for selecting an
element with some of these properties, it is enough to compute the
fixed point algebra (as explained above), which will be isomorphic
to a real Clifford algebra $Cl^{r,s}$ for some $r$ and
$s$\footnote{Notice that on $Cl^{r,s}$ there is not a Real
involution anymore and both $r$ and $s$ depend on the Real
involution defined in $~^p\IC l^{n,0}$ or equivalently in
$Cl^{n,0}.$}. Then we just need to compute the natural even and
odd part of $Cl^{r,s}$ as a Clifford algebra as explained in C.1.




\section{Equivariant KK-Theory }

In this appendix we will describe the pertinent modifications to
the KK-theory bifunctor described earlier to define the
equivariant KK$_{G}$-theory, which turns out to be the appropiate
tool for the classification of D-branes in orbifold singularities.
A $C^{*}$-algebra $A$ is called a $G$-algebra if there is a
compact group $G$ acting on it by the automorphism group, i.e., by
a map $ \alpha :G \mapsto Aut(A) $. In this appendix all
$C^{*}$-algebras are required to be $G$-algebras.

The $G$-action is said to be continuous if $ \alpha :G \mapsto
Aut(A) $ is continuous. This definition is rephrased by requiring
that the induced map $G \times A \mapsto A:(g,a) \mapsto g(a)$ is
norm continuous, where $A$ is realized as an operator algebra with
the strong operator topology.

By the Hilbert $G$-module $\mathcal{H}_{B}$ we mean the Hilbert
$B$-module $\mathcal{H}_{B}$ together with a linear action of $G$,
such that:

\begin{itemize}

\item $g(xb)=g(x)g(b)  \qquad {\rm{for \; all}}  \quad g \in G,
\quad x \in \mathcal{H}_{B}, \quad b \in B$,

\item $(g(x),g(y))=g((x,y)) \qquad {\rm{for \; all}}  \quad g \in
G, \quad x,y \in \mathcal{H}_{B}$,

\end{itemize}

\noindent where $(x,y)$ is the $B$-valued inner product of $
\mathcal{H}_{B}$ as a Hilbert $B$-module. We have the additional
condition that this action be norm-continuous i.e the map $g
\mapsto \| (gx,gx) \|$, $x \in  \mathcal{H}_{B}$ is norm
continuous in the strong operator topology. An element $x \in
\mathcal{H}_{B}$ is said to be invariant if $ g(x)=x$ for all $ g
\in G$.

In $ \mathbf{B}(\mathcal{H}_{B})$ there is an induced natural
action as follows: If $F \in \mathbf{B}(\mathcal{H}_{B})$, then
$g(F)$ is defined as $(g(F))(x)=g(F(g^{-1}(x))), \quad g \in G,
\quad x \in \mathcal{H}_{B}$. This induced $G$-action is not norm
continuous in general and those $F$ for which this holds are
called $G$-continuous. They make up a $C^{*}$-subalgebra
 of $\mathbf{B}(\mathcal{H}_{B})$ which contains
 $ \mathcal{K} (\mathcal{H}_{B})$.

 Now, consider  all even Kasparov $G$-modules $( \mathcal{H}_{B},
 \phi, F)$, i.e. the set of even Kasparov modules such that:

 \begin{itemize}

\item $\mathcal{H}_{B}$ is a $G$-Hilbert $B$-module,

\item $ \phi: A \mapsto \mathbf{B}(\mathcal{H}_{B})$ is an
equivariant $ \ast$-homomorphism,

\item $F \in \mathbf{B}(\mathcal{H}_{B})$ is an invariant (in
particular, $G$-continuous) element.

\end{itemize}

 Then the equivariant KK$_{G}$-group, denoted $KK_{G}(A,B)$ is
defined as the set of equivalence classes of even Kasparov
$G$-modules under the equivalence relation $\sim _{oh}$ defined
exactly as in appendix A, but with the additional condition that
the operator $U \in \mathbf{B}(\mathcal{H}_{B})$ generating the $
\mathit{unitarily \ equivalence}$ relation be invariant under the
$G$-action.

The higher KK$_{G}$-groups $KK_{G}^{n}(A,B)$ are defined as
before, i.e $KK_{G}^{n}(A,B)=KK_{G}^{n}(A,B \otimes
\mathbb{C}l^{n})$ where $G$ acts trivially on $\mathbb{C}l^{n}$
and again they have periodicity mod 2.

The KK$_{G}$-groups share the analog of the property (\ref{kkbp})
of $KK$-groups:

\begin{equation}
\qquad
$$KK^{k}_{G}(X,Y)$$=$$KK^{k-n}_{G}(X \times \mathbb{R}^{n},Y)$$=$$KK^{k+n}_{G}(X,Y
\times \mathbb{R}^{n}),$$ \label{kkbpg}
\end{equation}

\noindent with the additional requirement that $G$ acts on
$\mathbb{R}^{n}$
 by means of the spinor representation, i.e. by a group
 homomorphism $G \mapsto Spin_{n}$.


 One of the most important properties shared by all versions of
 Kasparov K-theory is additivity:

 \begin{equation}
$$KK_{G}(A,B_{1} \oplus B_{2} \oplus \cdots \oplus B_{n})=KK_{G}(A,B_{1})
\oplus KK_{G}(A,B_{2}) \oplus \cdots \oplus KK_{G}(A,B_{n})$$.
 \label{additivity}
 \end{equation}

\noindent A similar expresion also holds for the first of the
entries in the KK$_{G}$-functor.

 We mention it here because it is particularly
 important for calculating the low energy effective gauge theory
 living in the worldvolume of an unstable D-brane system placed
 at an orbifold singularity.

\bibliography{KKRjuly}
\addcontentsline{toc}{section}{Bibliography}
\bibliographystyle{TitleAndArxiv}
\end{document}